\documentstyle[preprint]{aastex}

\begin{document}

\title{Physical Properties of Main-Belt Comet 176P/LINEAR
        \footnote{Some of the data presented herein were obtained at the Gemini
       Observatory, which is operated by the Association of Universities for
       Research in Astronomy, Inc., under a cooperative
       agreement with the NSF on behalf of the Gemini partnership.
       Additionally, some of the presented data were obtained at the W. M.
       Keck Observatory, which is operated as a scientific partnership
       among the California Institute of Technology, the University of
       California, and the National Aeronautics and Space Administration,
       and made possible by the generous financial support
       of the W.~M.\ Keck Foundation.
       We are grateful for access to Lulin Observatory which is supported by the National Science Council
       of Taiwan, the Ministry of Education of Taiwan, and Taiwan's National Central University.
       Some data presented herein were also obtained at European Southern Observatory facilities at La Silla
       under program ID 081.C-0822(A).
       This work is partially based on observations made with the
       William Herschel Telescope and Isaac Newton Telescope operated on the
       island of La Palma by the Isaac Newton Group in the Spanish Observatorio del Roque de los Muchachos of
       the Instituto de Astrofisica de Canarias (programs W/2009A/23 and I/2009B/11).
       }
}

\author{Henry H.\ Hsieh$^{a,b,e}$, Masateru Ishiguro$^c$, Pedro Lacerda$^{b,f}$, and David Jewitt$^d$}

\affil{$^a$ Institute for Astronomy, Univ.\ of Hawaii, 2680 Woodlawn Drive, Honolulu, HI 96822, USA\newline
       $^b$ Astrophysics Research Centre, Queen's University,
    Belfast, BT7 1NN, United Kingdom\newline
       $^c$ Dept.\ of Physics and Astronomy, Seoul National University,
    Seoul, South Korea\newline
       $^d$ Dept.\ of Earth and Space Sciences and Institute for Geophysics and Planetary Physics, Univ.\ of California, Los Angeles, 3713 Geology Building, Box 951567, Los Angeles, CA 90095, USA\newline
       $^e$ Hubble Fellow\newline
       $^f$ Michael West Fellow
}

\email{hsieh@ifa.hawaii.edu, ishiguro@astro.snu.ac.kr, p.lacerda@qub.ac.uk, jewitt@ucla.edu}

\slugcomment{Submitted, 2011-02-28; Accepted, 2011-05-03}

\begin{abstract}
We present a physical characterization of comet 176P/LINEAR,
the third member of the new class of main-belt comets, which exhibit cometary
activity but are dynamically indistinguishable from main-belt asteroids, to be discovered.
Observations show the object exhibiting a fan-shaped tail for at least one month in
late 2005, but then becoming inactive in early 2006.  During this active period,
we measure broadband colors of $B-V=0.63\pm0.02$, $V-R=0.35\pm0.02$, and
$R-I=0.31\pm0.04$.  Using data from when the object was observed to be inactive,
we derive best-fit IAU phase function parameters of $H=15.10\pm0.05$~mag and $G=0.15\pm0.10$,
and best-fit linear phase function parameters of $m(1,1,0)=15.35\pm0.05$~mag and
$\beta=0.038\pm0.005$~mag~deg$^{-1}$.  From this baseline phase function, we find that
176P exhibits a mean photometric excess of $\sim$30\% during its active
period, implying an approximate total coma dust mass of $M_d\sim(7.2\pm3.6)\times10^4$~kg.
From inactive data obtained in early 2007, we find a rotation period of $P_{rot}=22.23\pm0.01$~hr
and a peak-to-trough photometric range of $\Delta m\sim0.7$~mag.  Phasing our photometric
data from 176P's 2005 active period to this rotation period, we find that the nucleus exhibits
a significantly smaller photometric range than in 2007 that cannot be accounted for by coma damping effects,
and as such, are attributed by us to viewing geometry effects.  A detailed analysis of these
geometric effects showed that 176P is likely to be a highly elongated object with an axis ratio
of $1.8<b/a<2.1$, an orbital obliquity of $\varepsilon\sim60\degr$, and a solstice position
at a true anomaly of $\nu_o=20\pm20\degr$.  Numerical modeling of 176P's dust emission found that
its activity can only be reproduced by asymmetric dust emission, such as a cometary jet.  We find
plausible fits to our observations using models assuming $\sim$10~$\mu$m dust particles
continuously emitted over the period during which 176P was observed to be active,
and a jet direction of $180\degr\lesssim\alpha_{jet}\lesssim120\degr$ and $\delta_{jet}\approx-60\degr$.
We do not find good fits to our observations using models of impulsive dust emission,
{\it i.e.}, what would be expected if 176P's activity was
an ejecta cloud resulting from an impact into non-volatile asteroid regolith.
Since for a rotating body, the time-averaged direction of a non-equatorial jet is equivalent to
the direction of the nearest rotation pole, we find an equivalent orbital obliquity
of $50\degr\lesssim\varepsilon\lesssim75\degr$, consistent with the results of our lightcurve analysis.
The results of our lightcurve analysis and dust modeling analysis are furthermore both
consistent with the seasonal heating hypothesis used to explain the modulation of 176P's activity.
Additional observations are highly encouraged to further characterize 176P's active behavior as
the object approaches perihelion on 2011 July 01.

\end{abstract}


\section{INTRODUCTION}
The cometary nature of 176P/LINEAR (also known as asteroid 118401; hereafter 176P)
was discovered on 2005 November 26 as part of the Hawaii Trails Project \citep[HTP;][]{hsi06c,hsi09a},
a targeted deep-imaging survey aimed at identifying cometary activity in main-belt asteroids.
Together with two other previously discovered comets --- 133P/Elst-Pizarro (hereafter, 133P)
and 238P/Read (hereafter, 238P) --- occupying orbits indistinguishable from
those of main-belt asteroids, the discovery of 176P's cometary nature led to the
designation of a new cometary class known as main-belt comets \citep[MBCs;][]{hsi06b}.

The cometary activity of each MBC is a strong indication of the presence of
extant ice (possibly preserved in subsurface layers) that has recently become exposed, perhaps
by collisions, and is now
sublimating.  Each MBC, however, has a Tisserand parameter (with respect to Jupiter) of
$T_J>3$, while classical comets have $T_J<3$ \citep{vag73,kre80}, suggesting that they
are unlikely to have been captured from the outer solar system and are most likely
native to the main belt \citep{hsi06b,jew09}.  The limited observational data used to discover
the known MBCs strongly implies that many more should exist \citep[$\gtrsim$100;][]{hsi09a},
indicating that present-day ice could be widespread in the main asteroid belt.
Recent detections of apparent water ice absorption at 3.1~$\mu$m in spectroscopic observations
of the surface of (24) Themis, the largest asteroid and namesake of
the Themis asteroid family, to which MBCs 133P and 176P also belong \citep{hsi09a}, appear to support this conclusion \citep{riv10,cam10},
though some believe the detected spectral feature could
instead be due to non-volatile materials \citep[{\it e.g.},][]{bec11}.  We note that this uncertainty does not explicitly
undermine the case for sublimating ice as the driver of MBC activity since, as stated above, that ice is believed to
largely reside in subsurface reservoirs, and so we would not necessarily expect it to be detectable
via reflectance spectroscopy of an apparently inactive object like (24) Themis.

\section{OBSERVATIONS}

The 2005 discovery of 176P's cometary nature was made using the 8~m Gemini North telescope on
Mauna Kea in Hawaii.  Following this discovery, confirmation and
characterization observations were made using Gemini North as well as the University of
Hawaii (UH) 2.2~m telescope, also on Mauna Kea.  Since then, numerous monitoring 
observations have been made using the UH 2.2~m and 10~m Keck I telescopes on Mauna Kea,
the 3.58~m New Technology Telescope (NTT) at the European Southern Observatory (ESO) at La Silla in Chile,
and the 2.5~m Isaac Newton Telescope (INT) and 4.2~m William Herschel Telescope (WHT) on La Palma in the Canary Islands.  The object
was also observed as part of the HTP using the Lulin 1.0~m telescope in Taiwan just one month
before its activity was discovered by Gemini, though no clear evidence of activity was seen in these
observations.  All reported observations (detailed further in Table~\ref{obslog}) were made in photometric conditions.

Observations with Gemini were made using the imaging mode of the Gemini Multi-Object Spectrograph
\citep[GMOS; image scale of $0\farcs146$ pixel$^{-1}$;][]{hoo04}, which uses Sloan Digital Sky Survey
$g'r'i'z'$ filters.
UH 2.2~m observations were made using either a Tektronix 2048$\times$2048 pixel CCD
with an image scale of $0\farcs219$ pixel$^{-1}$, or the Orthogonal Parallel Transfer Imaging Camera
\citep[OPTIC; $0\farcs14$ pixel$^{-1}$;][]{ton04}.
Keck observations were made using the imaging mode of the Low-Resolution Imaging Spectrograph
\citep[LRIS;][]{oke95}, which employs a Tektronix 2048$\times$2048
CCD ($0\farcs210$~pixel$^{-1}$).  
NTT observations were made using the ESO Faint Object Spectrograph and Camera \citep[EFOSC2;][]{buz84}
which employs a 2048$\times$2048~pixel Loral/Lesser CCD ($0\farcs24$~pixel$^{-1}$ using 2$\times$2 binning)
behind Bessel $BVR$ broadband filters.
INT observations were made using the Wide Field Camera
which consists of four thinned EEV 2048$\times$4096~pixel CCDs ($0\farcs333$~pixel$^{-1}$).
WHT observations were made using the Prime Focus Imaging Platform (PFIP)
which consists of two EEV 2048$\times$4096~pixel CCDs ($0\farcs24$~pixel$^{-1}$).
Lulin observations were made using a VersArray:1300B CCD \citep[$0\farcs516$~pixel$^{-1}$;][]{kin05}.
Except where otherwise specified, observations were made using
standard Kron-Cousins $BVRI$ broadband filters.

Standard bias subtraction and flat-field reduction
was performed on all images.  Flat fields were constructed from dithered images of
the twilight sky, except in the case of our Keck data which used flat fields
constructed from images of the illuminated interior of the telescope dome.
Photometry of \citet{lan92} standard stars and field stars
was obtained by measuring net fluxes (over sky background) within circular
apertures, with background sampled from surrounding circular annuli.
Object photometry was performed using circular apertures of different radii
(ranging from $2\farcs0$ to $5\farcs0$), but to avoid the contaminating
effects of the coma, background sky statistics were measured manually in
regions of blank sky near, but not adjacent, to the object.  Several (5--10) field
stars in the comet images were also measured and used to correct for minor extinction
variation during each night.

\section{RESULTS}

\subsection{Active Cometary Behavior}

For all nights on which 176P was observed, we combine individual $R$- or $r'$-band images
(aligned on the object's photocenter) into deep composite images
to assess the level of cometary activity present (Fig.~\ref{images_re70}).  Activity is unambiguously detected in
observations made of the object between UT 2005 November 26 and 2005 December 29.
In these observations, no coma is apparent, though a fan-shaped tail is clearly visible.
No activity in the form of a coma or tail is seen in the next available set of observations
obtained on 2006 February 03 nor in any monitoring observations made since that time.

Poor seeing during a precovery observation of 176P on 2005 October 24, just one month prior
to the first confirmed observations of activity, prevents conclusive determination of the
object's active nature at the time.  Photometric evidence (below), however, now suggests that the
object may have in fact been active.
Overall, 176P's behavior is consistent with that of an object that exhibits activity near perihelion
but not at other points in its orbit (Fig.~\ref{actv176p}).
In all observations when 176P is observed to be active, the tail is directed approximately
due east, roughly 20$^{\circ}$ south of the anti-solar direction.

We also measure $BVRI$ colors of 176P on several occasions during its 2005 active period
(Table~\ref{colors}), finding mean values of $B-V=0.63\pm0.02$, $V-R=0.35\pm0.02$, and
$R-I=0.31\pm0.04$, and no evidence of substantial variability.  This lack of variability
is unsurprising because, while no significant coma is observed, 176P may very well possess an
unresolved coma \citep[{\it cf}. 133P;][]{hsi10}.  If unresolved coma is present, not only will that coma not
exhibit color variability as the nucleus rotates, it will also dampen any color
variability that the underlying nucleus itself might exhibit.

\subsection{Phase Function\label{phasefunction}}

Characterization of an object's photometric dependence on changing solar phase angle, {\it i.e.}, its
phase function, permits data obtained at different observing geometries to be compared and combined,
while also giving insight as to the properties of the surface material of the object itself
\citep[{\it e.g.},][]{bow89}.  To compute 176P's phase function, we first consider only photometric data
obtained when 176P appears to be inactive ({\it i.e.}, from 2006 Feb 03 onwards; Table~\ref{obslog}).

On several nights, a significant length of time ($\gtrsim0.6$~hr)
elapses from the time of our first observation of the object to the time of the last observation,
during which the object could potentially undergo significant rotation.  We refer to these data as
``lightcurve observations'' (Table~\ref{ltcobs}).  Observations on other nights that do not
span a significant amount of time are referred to as ``snapshot observations''.  While on no occasion is
176P seen to reach both the maximum and minimum of its rotational lightcurve in a single
night of observations, we can nevertheless exploit these lightcurve observations to
obtain useful information regarding the midpoints of the rotational lightcurve on each night.

For computing 176P's phase function, we aim to use its mean brightness (averaged over a full rotation)
as a common reference point for data from different epochs.
It is obviously not possible to know {\it a priori} the rotational phase of the object at the time
of a given snapshot observation.  As such, in using photometry from snapshot observations to compute
176P's phase function, we assign a maximum uncertainty, $\sigma$, equal to half of the expected lightcurve range, $\Delta m$.
This allows for the range of possible rotational phases up to and including the extreme case where
the snapshot observation was obtained at either the object's maximum or minimum brightness.

Unlike in snapshot observations, the rotational phase of lightcurve observations can be constrained by
the observed brightness variation, $\delta m$, over the course of the observing period.  If $\delta m\approx\Delta m$, the
mean of the minimum and maximum brightness of the object during that set of lightcurve observations will be
close to the midpoint of the actual lightcurve.  We express these constraints by assigning uncertainties to
the midpoints of our lightcurve observations equal to $\sigma=(\Delta m-\delta m) / 2$.  The final uncertainties
listed in Table~\ref{ltcobs} reflect both these assigned uncertainties and the uncertainties in the computed
mean brightnesses themselves.

\begin{figure}
\plotone{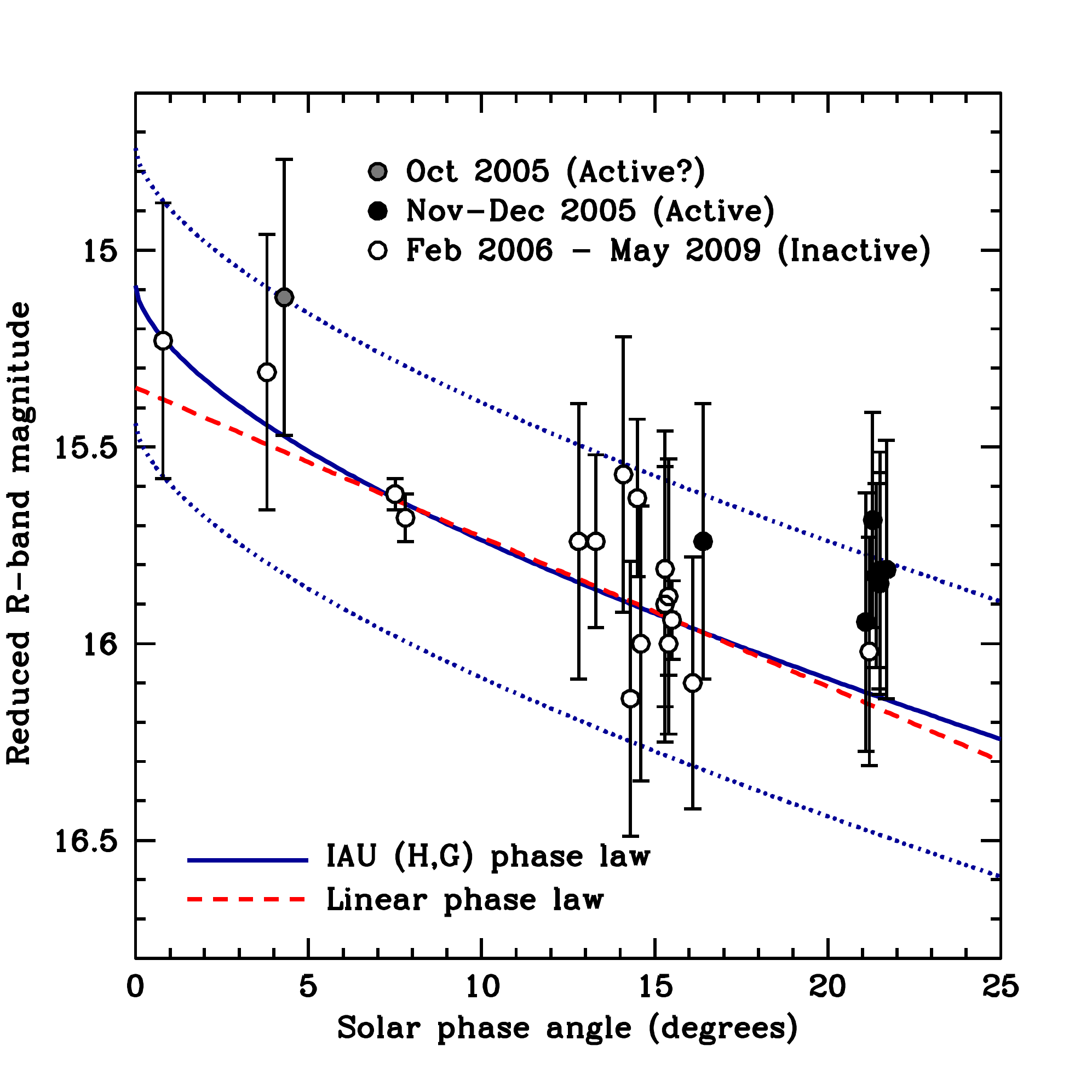}
\caption{\small Phase functions for 176P.  Points are estimated $R$-band magnitudes
  (normalized to heliocentric and geocentric distances of 1 AU; tabulated in
  Table~\ref{obslog}) at the mid-point of the full photometric range of the nucleus's
  rotational light curve. Solid circles denote photometry obtained while 176P was
  visibly active, while open circles denote photometry obtained while 176P appeared
  to be inactive. The dashed line represents a least-squares fit (excluding photometry
  points for which $\alpha<5^{\circ}$ where an opposition surge effect is expected) to a
  linear phase function where $m_R(1,1,0)=15.35\pm0.08$~mag and $\beta=0.038\pm0.008$~mag~deg$^{−1}$.
  The solid line represents an IAU ($H,G$) phase function fit where
  $H_R=15.09\pm0.05$~mag and $G_R=0.15\pm0.10$, while the dotted lines indicate the
  expected range of possible magnitude variations ($\sim0.35$~mag) due to the object's rotation.
}
\label{phaselaws}
\end{figure}

We then perform weighted fits of photometric data obtained from both lightcurve observations and snapshot
observations to the IAU $H,G$ phase function \citep{bow89}, finding best-fit parameters of $H=15.10\pm0.05$~mag
and $G=0.15\pm0.10$, which are in agreement with the parameters calculated in \citet{hsi09b} with fewer
data points and a less sophisticated weighting scheme.
We also fit our data to a linear phase function, omitting data points obtained at solar phase angles of
$\alpha\lesssim5^{\circ}$ at which opposition surge effects are expected, finding best-fit parameters of
$m(1,1,0)=15.35\pm0.05$~mag and $\beta=0.038\pm0.005$, again in agreement with \citet{hsi09b}.  These
solutions and the data used to compute them are plotted in Figure~\ref{phaselaws}.

For all of these calculations, we use $\Delta m=0.70$~mag based on the maximum photometric variation observed
in lightcurve observations (Table~\ref{ltcobs}).  We note, however, that our eventual best-fit phase function parameters are
not highly sensitive to the value chosen for $\Delta m$ as long as relative uncertainties for snapshot
and lightcurve observations are preserved.

\subsection{Photometric Confirmation of Activity\label{photactive}}

Once a phase function has been computed for 176P using inactive data, we can overlay photometric data for 176P
obtained when the object was active.  These data are plotted as solid circles in Figure~\ref{phaselaws} and can
be seen to be consistently brighter than the expected mean brightness of the object at those phase angles,
confirming the presence of excess flux attributable to cometary activity.  This type of photometric detection
of activity led to the discovery of activity in 95P/Chiron \citep{bus88,tho88,mee89,har90} and was also
used to detect unresolved coma for 133P \citep{hsi10} and to confirm the reactivation of 238P \citep{hsi11}.

For data taken when 176P was seen to be visibly active (2005 November 26 to 2005 December 29), we
find that the nucleus is, on average, $0.3\pm0.1$~mag brighter than expected, given the $H$ and $G$
phase function parameters we derive in \S\ref{phasefunction}.  Following \citet{hsi10}, we assume
that the discrepancy between the observed magnitude, $m_{obs}$, and the expected magnitude, $m_{exp}$, is due to dust
contamination, and that the albedos of the nucleus and dust are equal.
We then calculate that the scattering surface area of the dust coma (and any portion of the tail
also contained within the photometry aperture), $A_d$, is $\sim30$\% that of the inactive nucleus, $A_n$, using
\begin{equation}
{A_d\over A_n} = 10^{0.4(m_{exp}-m_{obs})}-1.
\label{ndratio}
\end{equation}
Using $A_n=\pi r_e^2=(1.3\pm0.2)\times10^7$~m$^2$, where $r_e=2.0\pm0.2$~km is the effective radius of 176P's nucleus \citep{hsi09b},
we therefore find $A_d=(0.4\pm0.2)\times10^7$~m$^2$.  Assuming a typical dust grain radius of 10~$\mu$m --- similar to dust
grain radii found for 133P \citep{hsi04} and consistent with dust modeling results for 176P (\S\ref{modeling}) --- and a bulk
grain density of $\rho_d=1300$~kg~m$^{-3}$ \citep{hsi04}, we find an approximate total coma dust mass of $M_d\sim(7.0\pm3.5)\times10^4$~kg,
comparable to coma dust masses found for 133P \citep[{\it cf}.][]{hsi10}.

\subsection{Rotation\label{rotation}}

\subsubsection{Period Determination\label{period}}

Given the minimal nucleus-obscuring coma observed during 176P's 2005 active phase, we
performed phase dispersion minimization analysis \citep[{\it e.g.},][]{ste78} on the
photometry data we obtained during that time in an attempt to ascertain the nucleus's
synodic rotation period, but were unable to identify a period that
would allow us to produce a convincing lightcurve.  At the time, we attributed this
failure to find a plausible rotation period to a probable extremely slow rotation
rate, which made it difficult to sample a significant portion of the rotational
lightcurve in a single night (where the object was visible for $\sim$4~hr in 2005 December),
and possible aliasing caused by a rotation period
close to the Earth's own daily 24-hr rotation period.

A more suitable data set for determining 176P's rotation period was obtained on two nights
in 2007 February and two additional nights in 2007 March.  In addition to affording longer
nightly visibility windows ($\sim$7~hr in February and $\sim$5~hr in March), the timing
of the observations over two consecutive months allowed us to avoid both aliasing effects
and complications due to significant changes in viewing geometry.  Applying the same phase dispersion
minimization techniques as before and phasing our data to candidate rotation periods to
assess plausibility, we find a likely rotation period of $P_{rot}=22.23\pm0.01$~hr (assuming the lightcurve is double-peaked)
and a peak-to-trough photometric range of $\Delta m\sim0.7$~mag (Fig.~\ref{ltcurve}; where
data for 2007 May 19 is overplotted to check for consistency).  We estimate the uncertainty on
the period by slowly varying the optimum period and re-phasing the data until we determine that
the lightcurve is noticeably out of phase.
We also find possible
but less likely candidate periods of $P_{rot}=22.57$~hr, $P_{rot}=22.83$~hr, and $P_{rot}=23.20$~hr,
all of which produce lightcurves with larger phase dispersions than for $P_{rot}=22.23$~hr.

The lightcurve we find using our most likely rotation period is incompletely sampled,
but otherwise appears convincing.  These results confirm our preliminary findings from
the analysis of our 2005 data that the rotation of the object is extremely slow, but
also show that the object is relatively elongated, with an axis ratio of $a/b>10^{0.4\Delta m}$,
or $a/b>1.9$.  Interestingly, the lightcurve also exhibits features such as a V-shaped minimum
and an inverted U-shaped maximum that are qualitatively like those of contact binaries
\citep[{\it e.g.},][]{she04,man07,lac08}, though the photometric range we observe for 176P
is not quite as large as is normally expected for such systems \citep[$\Delta m>0.9$~mag;][]{lac08}.

\subsubsection{2005 Photometry Revisited}

\begin{figure}
\plotone{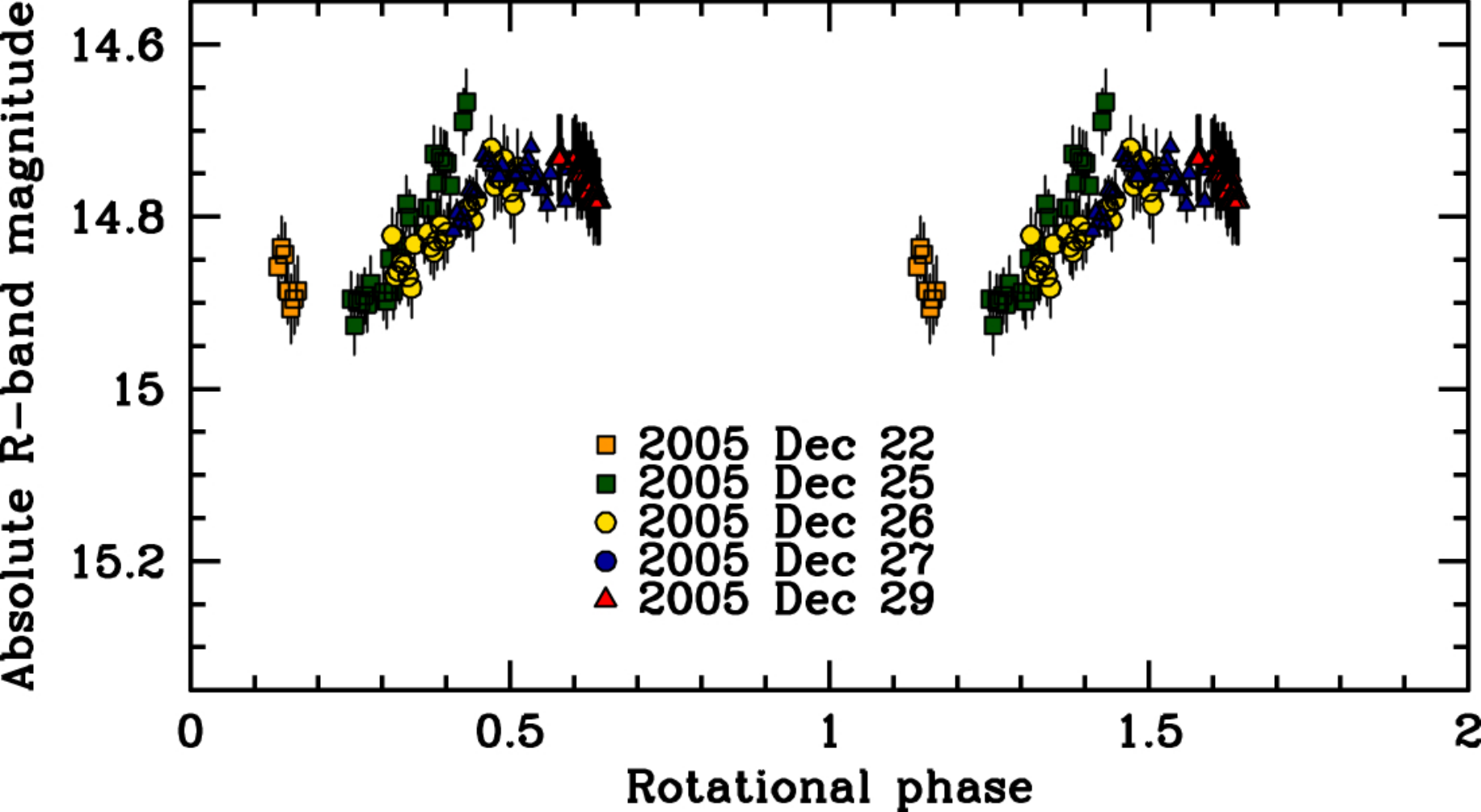}
\caption{\small Phase-angle-normalized, reduced-magnitude data ({\it i.e.}, normalized to $\alpha=0^{\circ}$ and $R=\Delta=1$~AU)
  for observations of 176P/LINEAR made between 2005 Dec 22 and 2005 Dec 29, phased to a rotation period of
  $P_{rot}=22.23$~hr.
}
\label{ltcurve_active}
\end{figure}

Having achieved a reasonable lightcurve using 2007 data (\S\ref{rotation}), we turn back to our 2005
data and attempt to phase them using the same rotation period (Fig.~\ref{ltcurve_active}).
When phased together, the data fit together far less well than the 2007 data, 
but the resemblance to the 2007 lightcurve is clear.  We attribute much of the scatter in the
data to the extreme sensitivity of photometry of active objects to seeing, where even small
changes in image quality from exposure to exposure can cause significantly different amounts
of dust contamination to be present in each photometry aperture.  We believe this effect to be the cause of
the slight brightness excess toward the second half of our 2005 December 25 observations
(Fig.~\ref{ltcurve_active}).  On 2005 December 24, we find an overall brightness excess for the comet
in the range of $\sim0.1-0.2$~mag (see Table~\ref{obslog}), but given that the comet was passing through a particularly dense star field on this
night, we attribute this excess to contamination from unseen faint field stars, and omit this data from
Figure~\ref{ltcurve_active} for clarity.
An active cometary object may of course be expected to occasionally exhibit short-lived outbursts, but
given the random (non-correlated with rotational phase) and short-lived nature of the observed photometric
anomalies, we find that they are adequately explained by fluctuations in the nightly seeing and, on occasion,
by faint field star contamination.

Significantly, the phased 2005 data appear to exhibit a far smaller photometric range than
the 2007 data,
showing a peak-to-trough variation of $\Delta m\sim0.2$~mag, down from $\Delta m\sim0.7$~mag in 2007.
As the observed photometric range of a rotating nucleus can be damped by a superimposed coma of constant brightness
(determined to be present for 176P in \S\ref{photactive}), before evaluating the physical implications of this change in
observed photometric range,
we must first determine how much of this effect could be due to coma damping.

Noting that the measured flux from an active comet, $F_{obs}$, consists of the sum of the fluxes
from the nucleus and dust contained within the photometry aperture, we can write
\begin{equation}
10^{0.4\Delta m_{obs}} = {F_{obs,max}\over F_{obs,min}} = \left({F_{n,max}+F_d\over F_{n,min}+F_d}\right)
\label{fluxobs}
\end{equation}
where $\Delta m_{obs}$ is the observed photometric range of the lightcurve, $F_{obs,max}$ and $F_{obs,min}$ are the maximum
and minimum fluxes, respectively, observed for the active nucleus, $F_{n,max}$ and $F_{n,min}$ are the maximum and minimum
fluxes, respectively, for which the nucleus is responsible, and the flux due to dust, $F_d$, is assumed to be constant.

For a true nucleus axis ratio ({\it i.e.}, uncontaminated by coma) of $(a/b)_n$, nucleus fluxes can be expressed as follows:
\begin{equation}
F_{n,max}=\left({a\over b}\right)_n^{1/2} F_n
\label{fluxmax}
\end{equation}
and
\begin{equation}
F_{n,min}=\left({a\over b}\right)_n^{-1/2} F_n
\label{fluxmin}
\end{equation}
where $F_n$ is the average flux of the nucleus over one full rotation.

Then, substituting Equations~\ref{fluxmax} and \ref{fluxmin} into Equation~\ref{fluxobs}, we find
\begin{equation}
\left({a\over b}\right)_n + \left[{F_d \over F_n}\left({1-10^{0.4\Delta m_{obs}}}\right)\right] \left({a\over b}\right)_n^{1/2} - 10^{0.4\Delta m_{obs}} = 0
\label{trueaxisratio}
\end{equation}
which can then be solved for $(a/b)_n^{1/2}$ using standard techniques for solving quadratic polynomials,
where $F_d/F_n=A_d/A_n$ is given by Equation~\ref{ndratio}.
Taking the positive root of Equation~\ref{trueaxisratio} (since Equations~\ref{fluxmax} and \ref{fluxmin} must produce positive values),
we find $(a/b)_n=1.27$, meaning that even accounting for coma contamination, the photometric range of the bare nucleus in 2005 is $\Delta m_n=0.26$,
still much smaller than that observed in 2007.

\subsubsection{Implications for Pole Orientation and Physical Nature\label{rotimplications}}

The plausibility of the seasonal heating hypothesis as a mechanism for modulating MBC activity \citep[first proposed for 133P;][]{hsi04},
depends crucially on each object having a pole orientation compatible with its activity profile.  Pole orientations are currently unknown
for all MBCs, though some constraints placed on 133P's pole orientation by \citet{tot06} are thus far consistent with seasonal modulation of
that MBC's activity.

\begin{figure}
\plotone{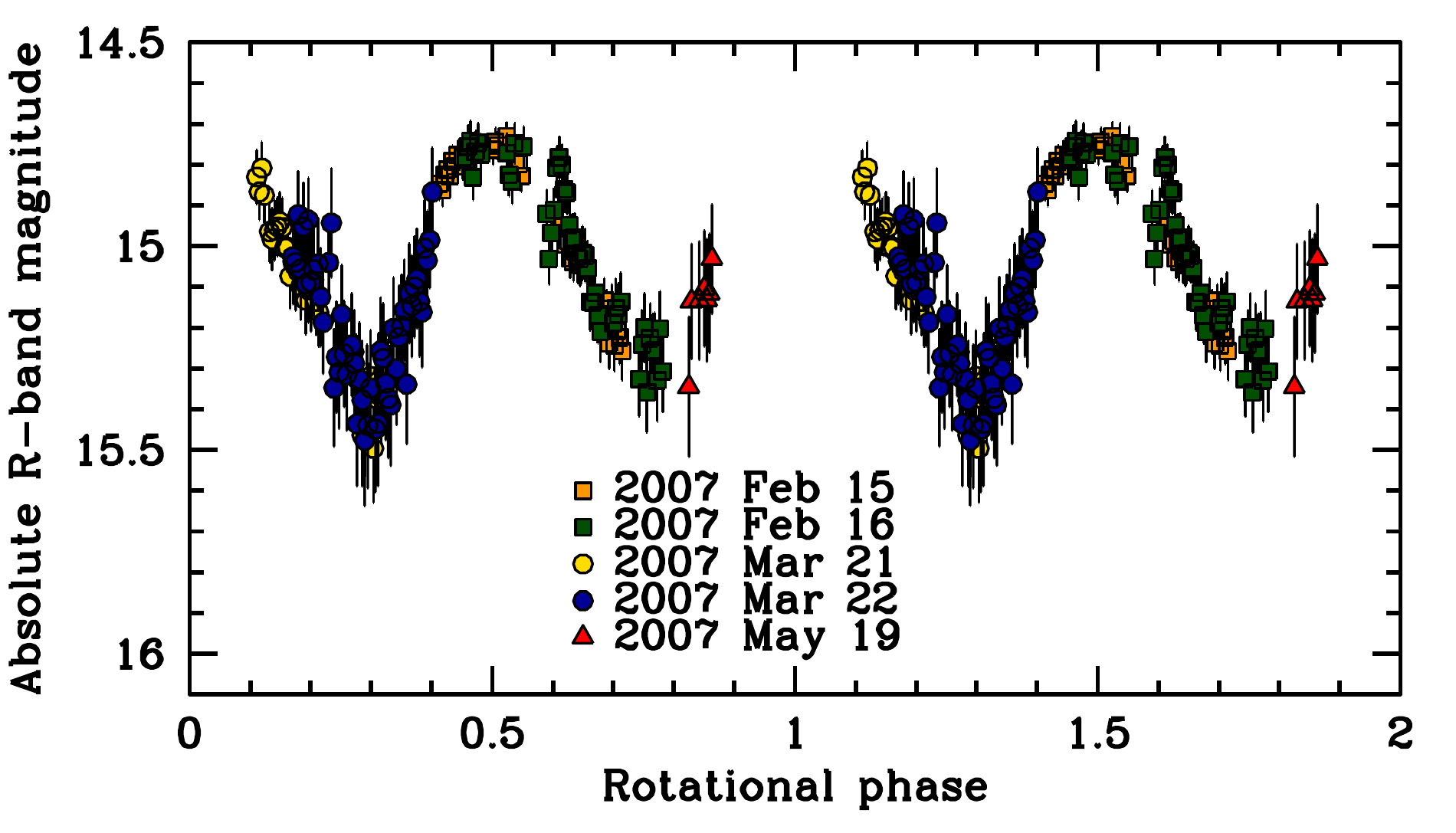}
\caption{\small Phase-angle-normalized, reduced-magnitude data ({\it i.e.}, normalized to $\alpha=0^{\circ}$ and $R=\Delta=1$~AU)
  for observations of 176P/LINEAR made between 2007 February and May, phased to a rotation period of
  $P_{rot}=22.23$~hr.
}
\label{ltcurve}
\end{figure}

In the case of 176P, we can use the rotational lightcurves shown in Figures~\ref{ltcurve} and \ref{ltcurve_active} to place constraints
on its pole orientation.
These figures show that the photometric range of 176P's lightcurve changed
appreciably between 2005 ($\Delta m\sim0.2$ mag) and 2007 ($\Delta m\sim0.7$
mag), which suggests that the object has significant obliquity.  The available data
(incomplete lightcurves at two epochs) are not sufficient to uniquely
determine the spin pole direction of 176P. For example, we are unable to break
the degeneracy between north and south pole.  However, we can still use those
data to obtain useful constraints on the rotational pole of 176P.

In particular, we are interested in constraining the shape and obliquity of 176P
and verifying if the data are consistent with the seasonal hypothesis for MBC
activity. According to the seasonal hypothesis, 176P should have non-zero
obliquity and be active close to one of its solstices. We thus elect to
parameterize the orientation of the spin pole of 176P by its obliquity,
$\varepsilon$, and by the true anomaly of the solstice, $\nu_0$, which we
assume to be the northern hemisphere summer solstice. For instance, if 176P
were at solstice exactly at perihelion, then $\nu_0=0$\degr. We assume the
already low orbital inclination ($i=0.23$\degr) of 176P to be exactly zero for
the purpose of this analysis. 
 
\begin{figure}
\plotone{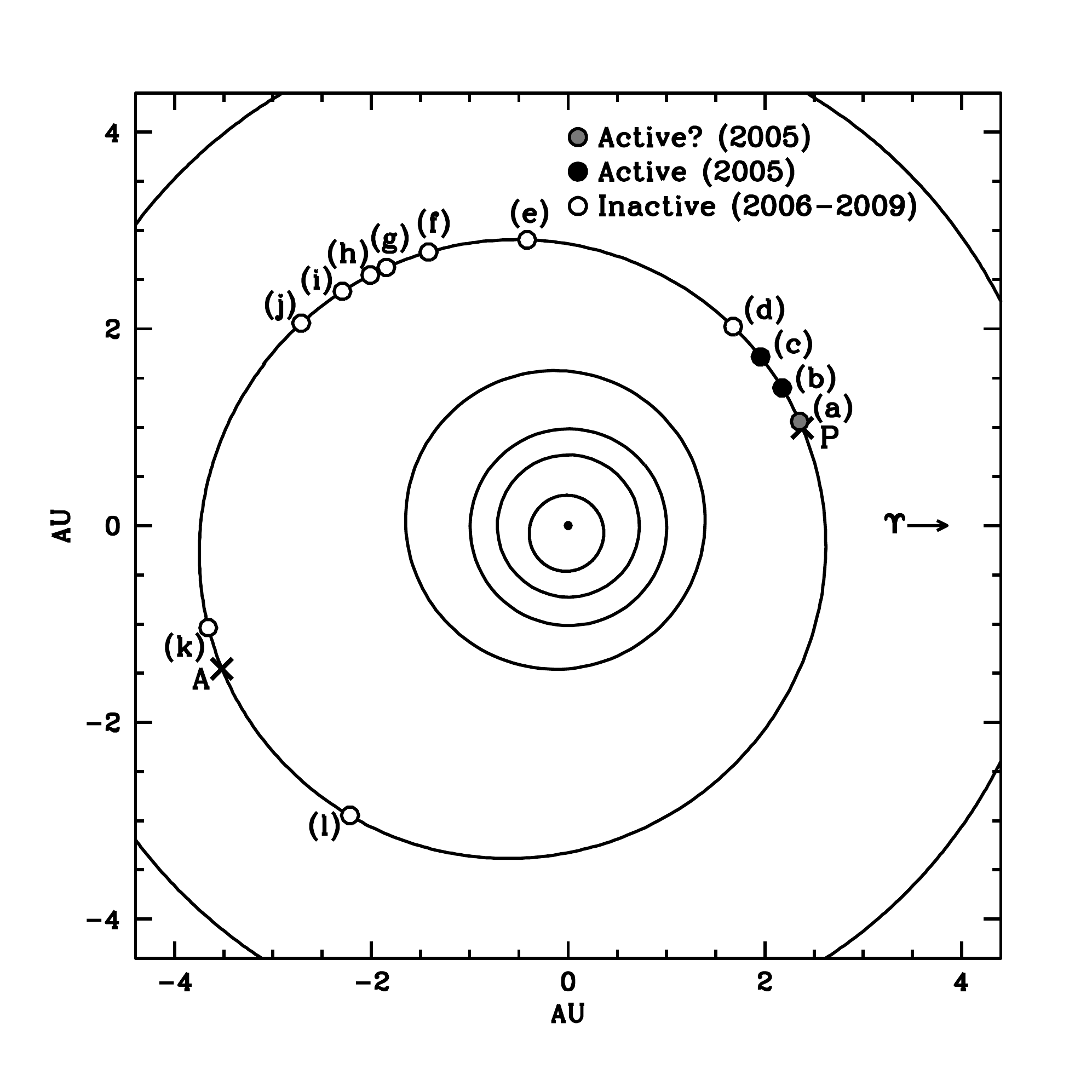}
\caption{\small Orbital positon plot of active and inactive phases of 176P
  detailed in Table~\ref{obslog}.  The Sun is shown at the center as a solid dot,
  with the orbits of Mercury, Venus, Earth, Mars, 176P, and Jupiter (from the
  centre of the plot outwards) are shown as black lines.
  Solid circles mark positions where
  176P was observed to be active, while open circles mark positions where
  176P was observed to be inactive.  Perihelion (P) and aphelion (A)
  positions are also marked with crosses.  References: (a) 2005 Oct 24 \citep{hsi09a},
  (b) 2005 Nov 26 \citep{hsi06b};  (c) 2005 Dec 22-29 \citep{hsi06b};
  (d) 2006 Feb 03-08; (e) 2006 Aug 31 - Sep 02; (f) 2006 Dec 11-18;
  (g) 2007 Jan 27; (h) 2007 Feb 15-16; (i) 2007 Mar 21-22;
  (j) 2007 May 19; (k) 2008 Jun 29 - Jul 01; (l) 2009 May 03,
  where (d)-(l) are from this work.
}
\label{actv176p}
\end{figure}

To investigate the lightcurve behavior of 176P we employ the simulations
described in \citet{lac07}. These simulations place triaxial
ellipsoids (semi-axes $a\geq b\geq c$) at pre-selected illumination and viewing
angles and register the integrated reflected flux as a function of rotational
phase (assuming simple principal axis rotation about the ellipsoids' short axes)
to extract modeled lightcurves.  In addition to observing
geometry (completely defined by $\varepsilon$, $\nu_0$, and the phase angle
$\alpha$), the model takes as parameters the axis ratios ($0 < c/a \leq b/a
\leq 1$) of the triaxial ellipsoid and the scattering law. For simplicity, we
use prolate ellipsoids ($c/a=b/a$) and a Lommel-Seeliger ``lunar'' scattering
function which has no free parameters and is appropriate for simulating the low
albedo \citep[$p_R=0.06\pm0.02$;][]{hsi09b} surface of 176P. In what
follows, we compare the model lightcurves with the data (Figs.~\ref{ltcurve} and \ref{ltcurve_active}) from
2005 (chiefly taken at point (c) in Fig.~\ref{actv176p}) and 2007 (taken at points (h), (i)
and (j) in Fig.~\ref{actv176p}).

Due to the underconstrained nature of this problem given the available data,
we only consider two limiting scenarios, one in which we simply assume that
the lightcurve amplitude observed in 2007 represents the object's maximum range,
placing it at equinox during those observations, and a second scenario that is
physically motivated where we assume that the object was receiving maximal heating
at the time when our 2005 observations showed it to be active, placing solstice
at perihelion.

First, we consider the possibility that 176P was close to equinox during the
2007 measurements, implying that the solstice occurs around true anomaly
$\nu_0=21$\degr, {\it i.e.}, just after the 2005 observations. In that case we can
use the photometric range in 2007 to obtain a direct estimate of the $b/a$ axis
ratio from the relation $\Delta m=-2.5\log(b/a)$. Using $\Delta m\sim0.7$
mag, we find $b/a\sim0.52$.

\begin{figure} \centering
\includegraphics[width=0.8\textwidth]{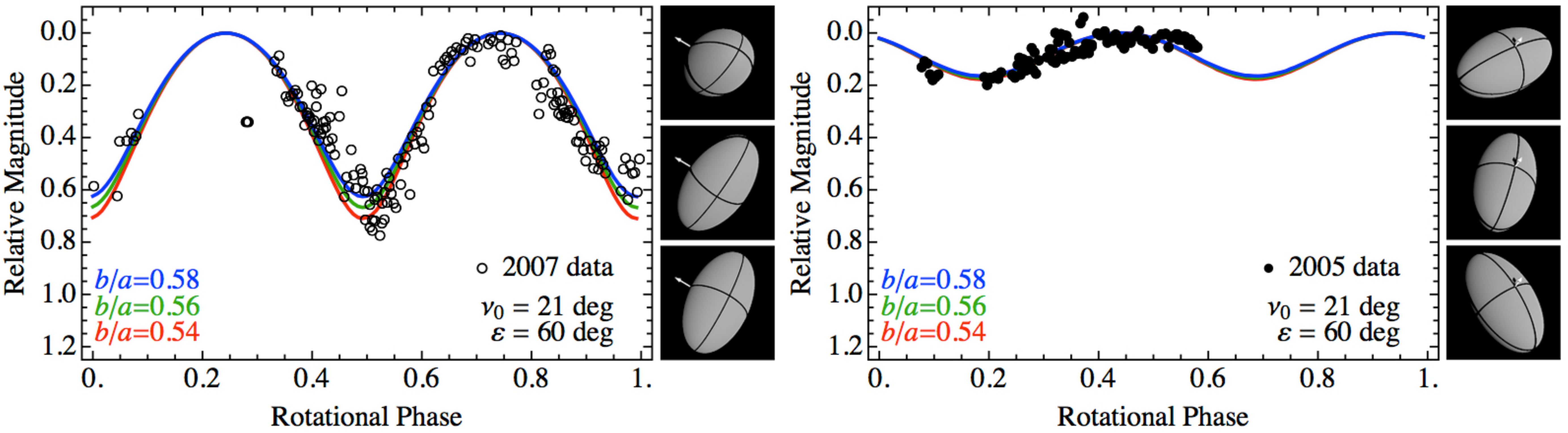} 
\caption{Simulated lightcurves (solid lines) based on triaxial ellipsoids
(shown to the right of each lightcurve plot) with
axis ratios $b/a=0.54$ (red), $b/a=0.56$ (green) and $b/a=0.58$ (blue) located
at the orbital configuration of 176P during the 2007 (left panel) and 2005
(right panel) observations. The simulations assume obliquity
$\varepsilon=60$\degr, a solstice position $\nu_0=21$\degr. The 2007 and 2005
data are overplotted as open and filled circles, respectively.
\label{P1}}
\end{figure}

Figure \ref{P1} (left panel) shows the 2007
lightcurve data overplotted on three simulations spanning $0.54\leq
b/a\leq0.58$. The reason why the best-fit model shapes are slightly less
elongated than the purely geometric estimate, $b/a=0.52$, is that limb darkening
effects contribute to increase the photometric variability beyond the geometric
expectation. Each model was calculated using the mean phase angle and true
anomaly of the 2007 observations. Under these conditions, the obliquity is
approximately determined by the 2005 ``solstice'' data.  An obliquity of
$\varepsilon=0$\degr\ would result in a nearly constant photometric range
throughout the orbit, whereas $\varepsilon=90$\degr\ would imply negligible
photometric variation in 2005. We find that an obliquity
$\varepsilon\sim60$\degr\ for an object with an axis ratio $b/a\sim0.56\pm0.03$
fits the 2005 data well (Fig.\ \ref{P1}, right panel).

\begin{figure}
\centering
\includegraphics[width=0.8\textwidth]{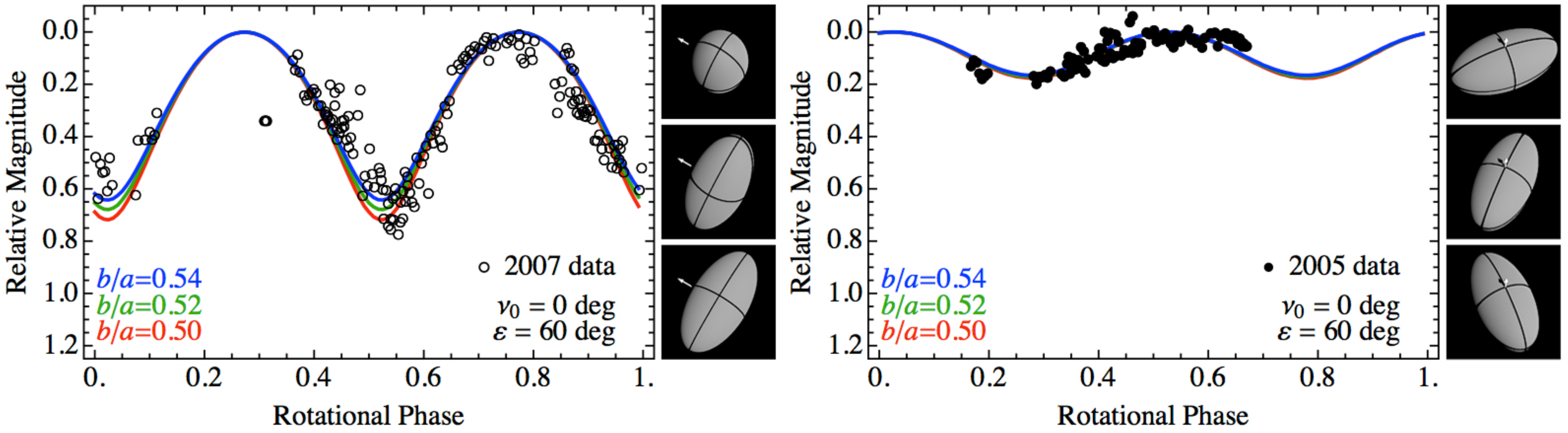}
\caption{Same as Fig.\ \ref{P1} but for solstice position $\nu_0=0$\degr\ and
ellipsoids with axis ratios $b/a=0.50$ (red), $b/a=0.52$ (green), and
$b/a=0.54$ (blue).\label{P3}}
\end{figure}

We now consider a second scenario in which solstice occurs at perihelion, $\nu_0=0$\degr, just
before the period of activity in 2005. We reason that if the activity is
powered by the seasonal heating of sub-surface ice, then it should begin at or
shortly after the time of maximum localized heating, {\it i.e.}, the solstice. If
$\nu_0=0$\degr, then equinox takes place at true anomaly $\nu=90$\degr, or
about 20\degr\ before the 2007 data were taken. In this scenario, 176P must be
more elongated than determined above. In Figure~\ref{P3} we show this to be the
case: to simultaneously fit the 2005 and 2007 data, for an obliquity of
$\varepsilon=60$\degr, an axis ratio $b/a\sim0.52$ is required.

\begin{figure}
\centering
\includegraphics[width=0.8\textwidth]{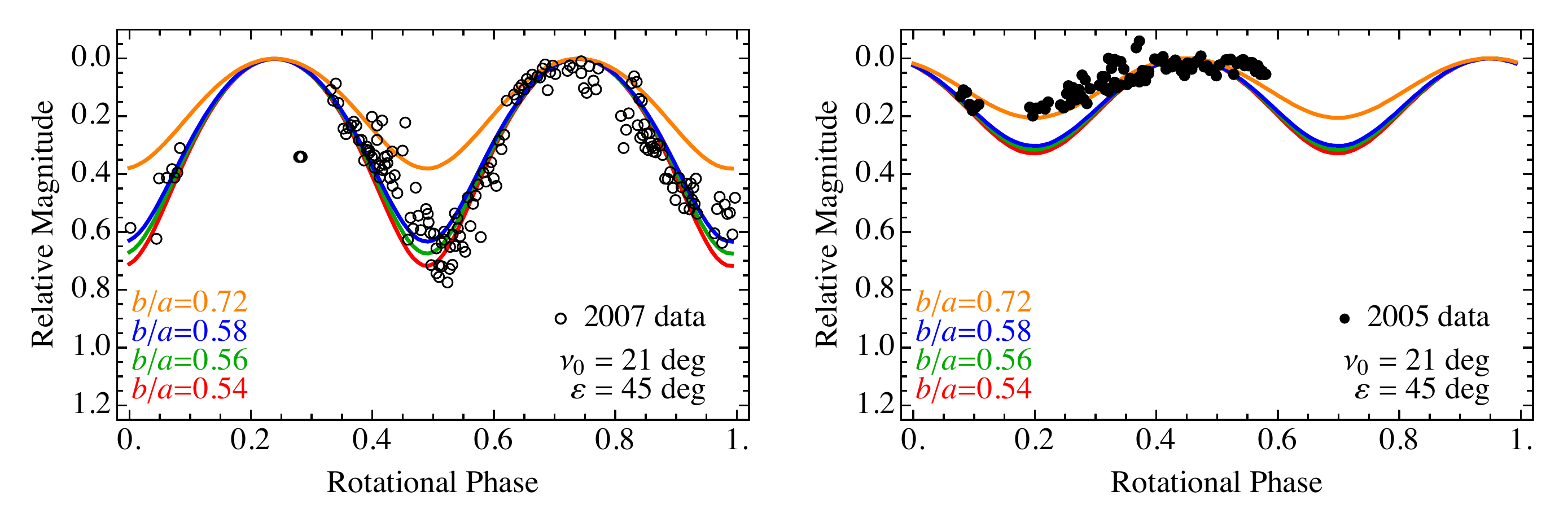}
\caption{Same as Fig.\ \ref{P1} but for obliquity $\varepsilon=45$\degr and
an ellipsoid with axis ratio $b/a=0.72$ (solid orange
line).\label{P2}}
\end{figure}

\begin{figure}
\centering
\includegraphics[width=0.8\textwidth]{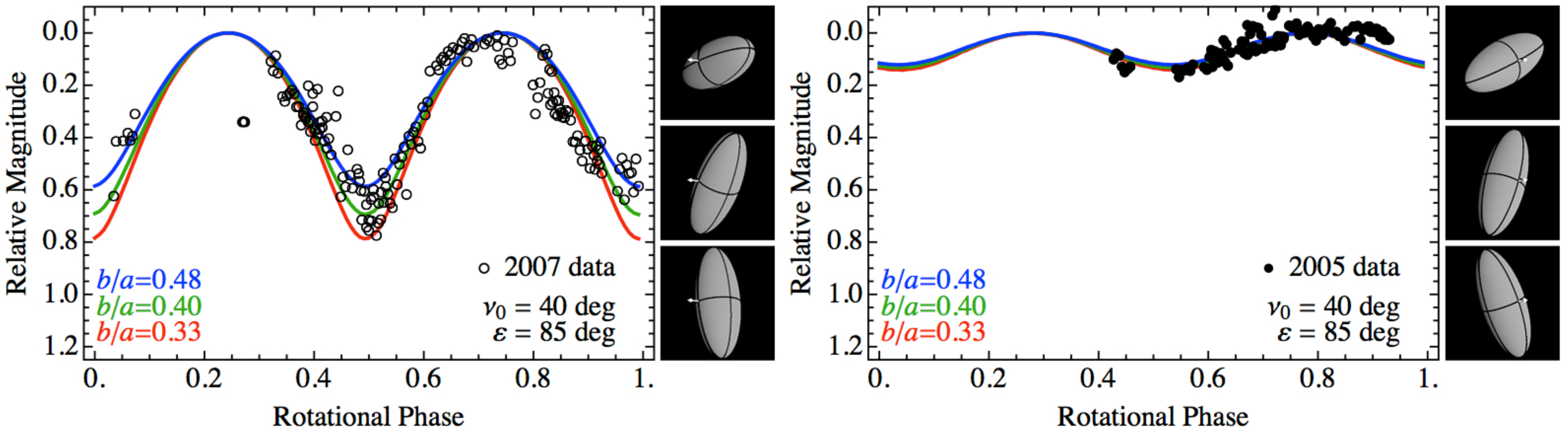}
\caption{Same as Fig.\ \ref{P1} but for obliquity $\varepsilon=85$\degr,
solstice position $\nu_0=40$\degr\ and ellipsoids with axis ratios $b/a=0.33$
(red), $b/a=0.40$ (green), and $b/a=0.48$ (blue).\label{P5}}
\end{figure}

In both scenarios, we note that models using significantly smaller obliquities ({\it e.g.},
$\varepsilon=45$\degr) cannot simultaneously fit the 2005 and 2007 ({\it e.g.}, Fig.~\ref{P2}).
Higher obliquities are possible, but require a more
elongated nucleus shape for 176P and a later solstice location along the orbit.
Figure \ref{P5} shows the result of simulations for solstice at
$\nu_0=40$\degr, obliquity $\varepsilon=85$\degr, and shapes $0.33\leq b/a\leq
0.48$. However, objects with such extreme axis ratios (3:1, 2.5:1 and 2.1:1 in
Fig.~\ref{P5}) are hydrostatically unstable \citep{jea19} and require
significant material strength to retain their shape. Alternatively, such
extreme configurations may be explained by contact binaries \citep{lac07}, but
our current data set is insufficient for confirming or ruling out this possibility.
Solstice positions significantly outside the interval $0\degr\leq\nu_0\leq 40\degr$\
(or equivalently, $180\degr\leq\nu_0\leq220\degr$, due to the fact that ``summer''
and ``winter'' solstices are indistinguishable in this analysis)
are untenable as the 2005 data with its smaller $\Delta m$ would fall as close or closer
to the equinox than the 2007 data with its larger $\Delta m$, when in fact
$\Delta m$ should reach a maximum at equinox.

In conclusion, our data strongly suggest that 176P is highly elongated, close
to a 2:1 axis ratio, and has significant obliquity, close to
$\varepsilon=60$\degr. The solstice positions that best fit the data are
located close to the active portion of the orbit, around $\nu_0=20\pm20$\degr,
placing the other solstice at $\nu_{180}=200\pm20$\degr, and equinoxes at $\nu_{90}=110\pm20$\degr
and $\nu_{270}=290\pm20$\degr.
These constraints on the obliquity and
solstice position of 176P are consistent with the seasonal hypothesis to explain the activity of
MBCs.

\section{DUST MODELING}\label{modeling}

In order to place quantitative constraints on 176P's dust emission, we generate a series
of numerical models for the dust and attempt to match them to our observations.  Due to
the limited observational data available and the resulting underconstrained nature of this
dust modeling effort, we recognize from the outset that it will not be possible to achieve
an exact model of 176P's activity.  As such, we purposefully formulate our modeling strategy
to simply achieve constraints on particular key properties such as 
grain sizes, ejection velocities, and the temporal behavior of the dust emission.

\begin{figure}
\plotone{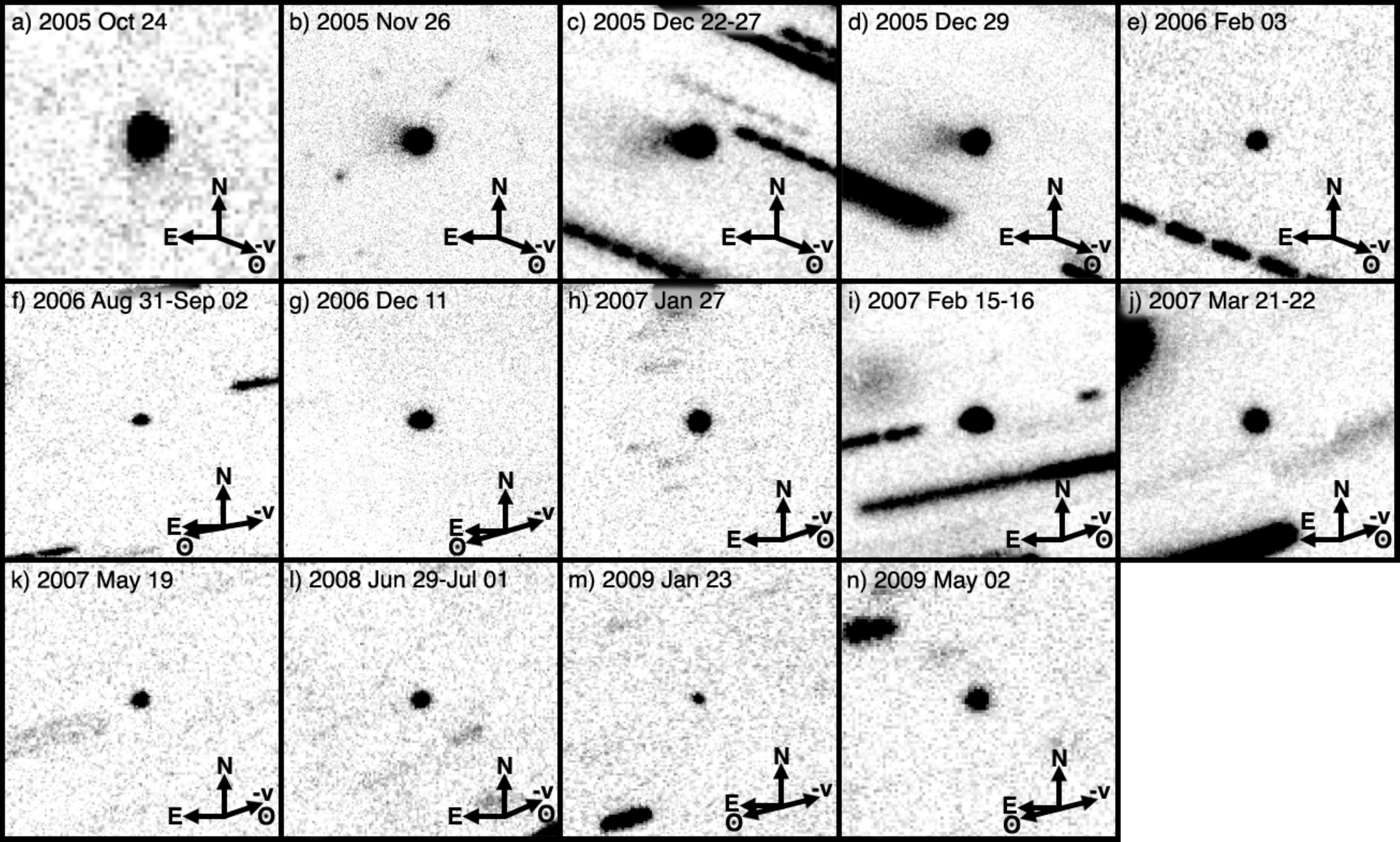}
\caption{\small Composite images of 176P from $R$-band images taken during observations detailed in
Table~\ref{obslog}.  Each image is $0\farcm5\times0\farcm5$ with 176P at the centre, with arrows
indicating north (N), east (E), the negative heliocentric velocity vector ($-v$), and the direction
towards the Sun ($\odot$).
Images shown comprise (a) 3000~s of exposure time on the Lulin 1.0~m telescope,
(b) 240~s on the 8~m Gemini North telescope, (c) 37500~s on the University of Hawaii 2.2~m telescope,
(d) 2520~s on Gemini North, (e) 3600~s on the UH 2.2~m, (f) 2400~s on the UH 2.2~m,
(g) 9000~s on the UH 2.2~m, (h) 240~s on the 10~m Keck I telescope, (i) 24600~s on the UH 2.2~m,
(j) 25800~s on the UH 2.2~m, (k) 3000~s on the UH 2.2~m, (l) 1440~s on the 3.54~m New Technology Telescope,
(m) 240~s on the 4.2~m William Herschel Telescope, and (n) 600~s on the 2.5~m Isaac Newton Telescope.
}
\label{images_re70}
\end{figure}

Standard dust modeling \citep[{\it e.g.},][]{fin68} typically makes use of syndyne curves,
which are lines representing the positions of particles of constant sizes ejected at different times
where ejection velocities are assumed to be zero, and synchrone curves, which are lines representing
the positions of particles of different sizes ejected at the same time with zero velocity.  In the case of 176P,
an extremely low inclination means that all synchrone and syndyne curves actually overlap with one
another in the object's orbit plane, making it very difficult to use this type of analysis to study
176P's dust emission.  We note, however, that the orientation of 176P's dust tail does not coincide
with its orbit plane, and thus, does not actually coincide with
{\it any} syndyne or synchrone curves (Fig.~\ref{images_re70}b,c,d), and therefore
we must consider asymmetric dust emission (as in a jet), similar to that considered in our previous
analysis of 238P \citep{hsi09c}.  The methods used in the following analysis are the same as
those used in that previous work except where otherwise specified. 

\begin{figure}
\plotone{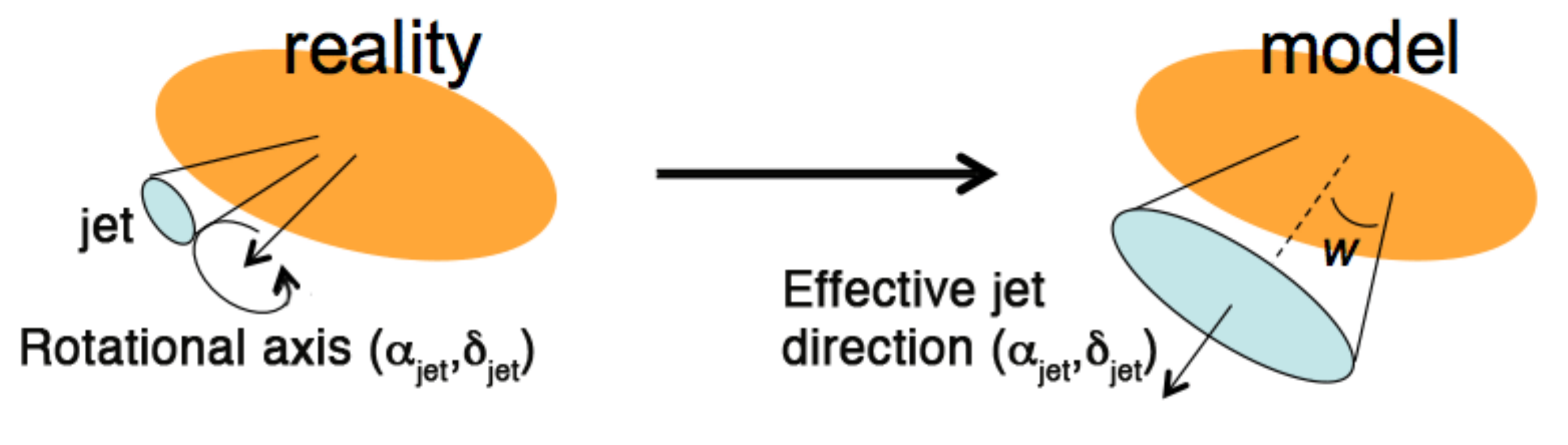}
\caption{\small Schematic diagram showing how a near-polar jet of directed ejected material
can be approximated by a wider jet with an effective orientation equivalent
to the direction of the rotational pole itself.
}
\label{jetmodel}
\end{figure}

Assuming jet-driven dust emission, we assume that the central axis of the cone points in a particular direction
in the inertial frame ({\it i.e.}, toward right ascension $\alpha_{jet}$, and declination $\delta_{jet}$). 
For a rotating body and a non-equatorial jet, it should be noted that the effective time-averaged jet position
in this model actually simply corresponds to the nearest rotational pole, where the derived jet width will be
somewhat larger than the actual jet width due to the sweeping motion caused by the body's rotation ({\it cf}.\
Fig.~\ref{jetmodel}).  For simplicity, we assume that Earth-bound observers are situated outside the cone
defined by the rotating jet, and that the strength of the jet does not vary with rotational phase.

The time-averaged jet is presumed to originate at the subsolar point
and have a half-opening angle of $w=45$\degr.  We furthermore assume that dust particles are released homogeneously from
a spherical body and that emission only occurs on the day-time side of the body.  Dust particle sizes are
parameterized in standard fashion where $\beta$ denotes the ratio of a particle's acceleration due to
solar radiation pressure to its acceleration due to gravity.  Terminal velocities,
$v_{ej}$, of ejected dust particles are given by
\begin{equation}
v_{ej}(r_h,\beta) = v_0\beta^{u_1}\left({r_h\over{\rm AU}}\right)^{-u_2}
\label{ejecvel}
\end{equation}
where $r_h$ is the heliocentric distance, $v_0$ is the reference ejection velocity in m~s$^{-1}$
\citep[assumed here to be $v_0=25$~m~s$^{-1}$ based on our work with 238P;][]{hsi09c} of particles
with $\beta=1$ at $r_h=1$~AU, and $u_1$ and $u_2$ (assumed here to be $u_1=u_2=0.50$) are the power indices of the reference ejection velocity
dependence on $\beta$ and $r_h$.  We use an exponential size distribution with an index of $q=-3.5$, minimum $\beta$
value of $\beta_{min}=9\times10^{-2}$, and maximum $\beta$ value of $\beta_{max}=10^{-1}$, and an
exponential dust production rate dependence on heliocentric distance with an index of $k=3$.  Dust emission is
assumed to begin 1 year prior to perihelion passage and is terminated on 2006 January 1 (three days after 176P was
last observed to be active).  We assume such an early start date to allow time for the dust to evolve to a ``steady'' state, but
given the short dissipation timescales for micron-scale dust particles,
also note that most of the dust that is actually observed is likely to have been ejected far more recently.

\begin{figure}
\plotone{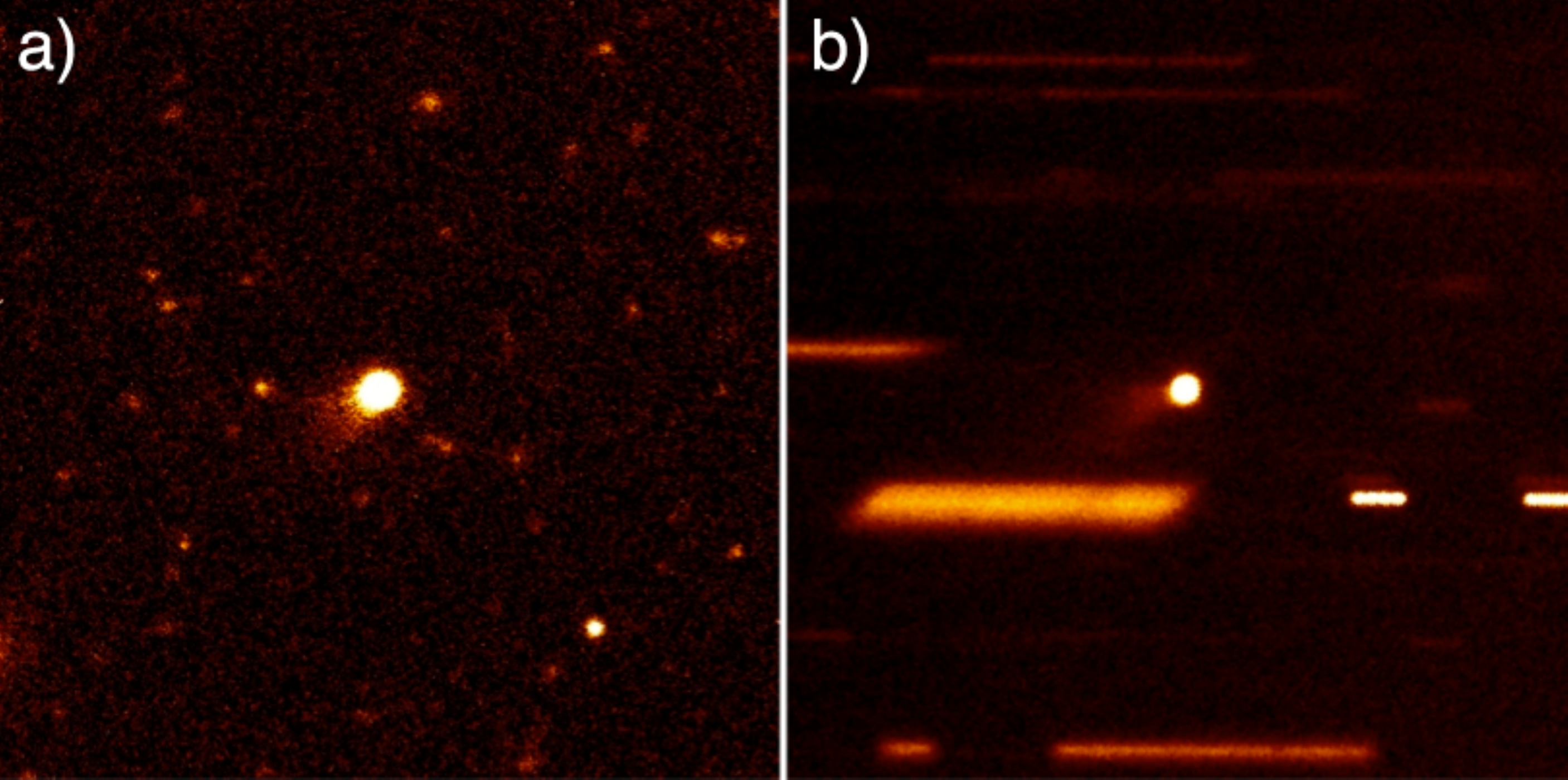}
\caption{\small Observations of 176P (obtained with Gemini North on (a) 2005 Nov 26 and (b) 2005 Dec 29) used to constrain numerical
models (\S\ref{modeling}), where images are shown in the ecliptic coordinate
system such that the orbital plane of 176P is effectively horizontal and where radiation pressure pushes dust particles
to the left.
}
\label{model_gemini}
\end{figure}

\begin{figure}
\plotone{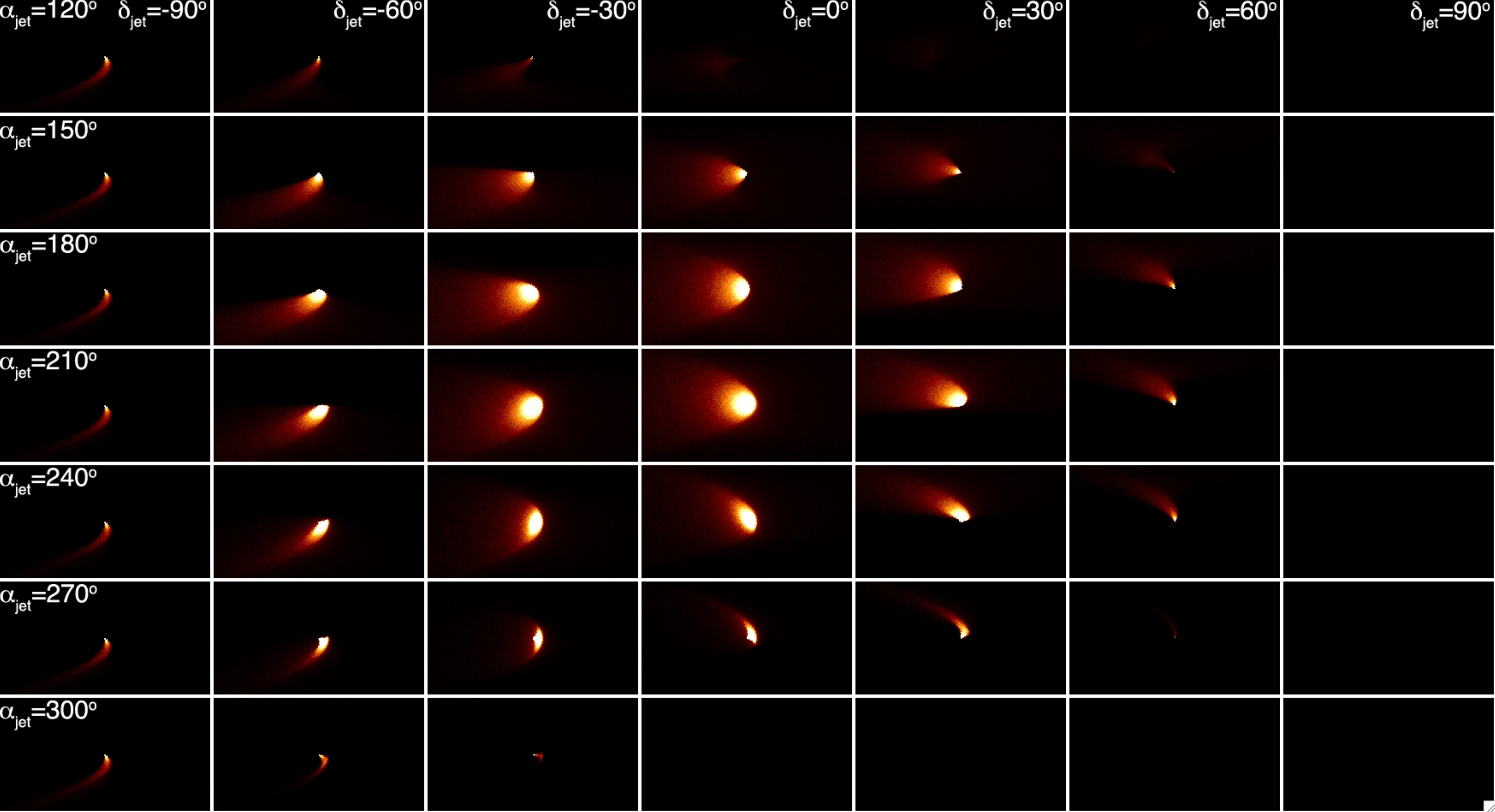}
\caption{\small Images of dust ejection models for 2005 Nov 26 for $10^{-3}<\beta<10^{-1}$ and
different jet directions as labeled, where $\alpha_{\rm jet}$ is constant for each row of models
and $\delta_{\rm jet}$ is constant for each column of models.
In all panels, the source of emission ({\it i.e.}, the nucleus) is at the center of each image.
}
\label{model_20051126}
\end{figure}

\begin{figure}
\plotone{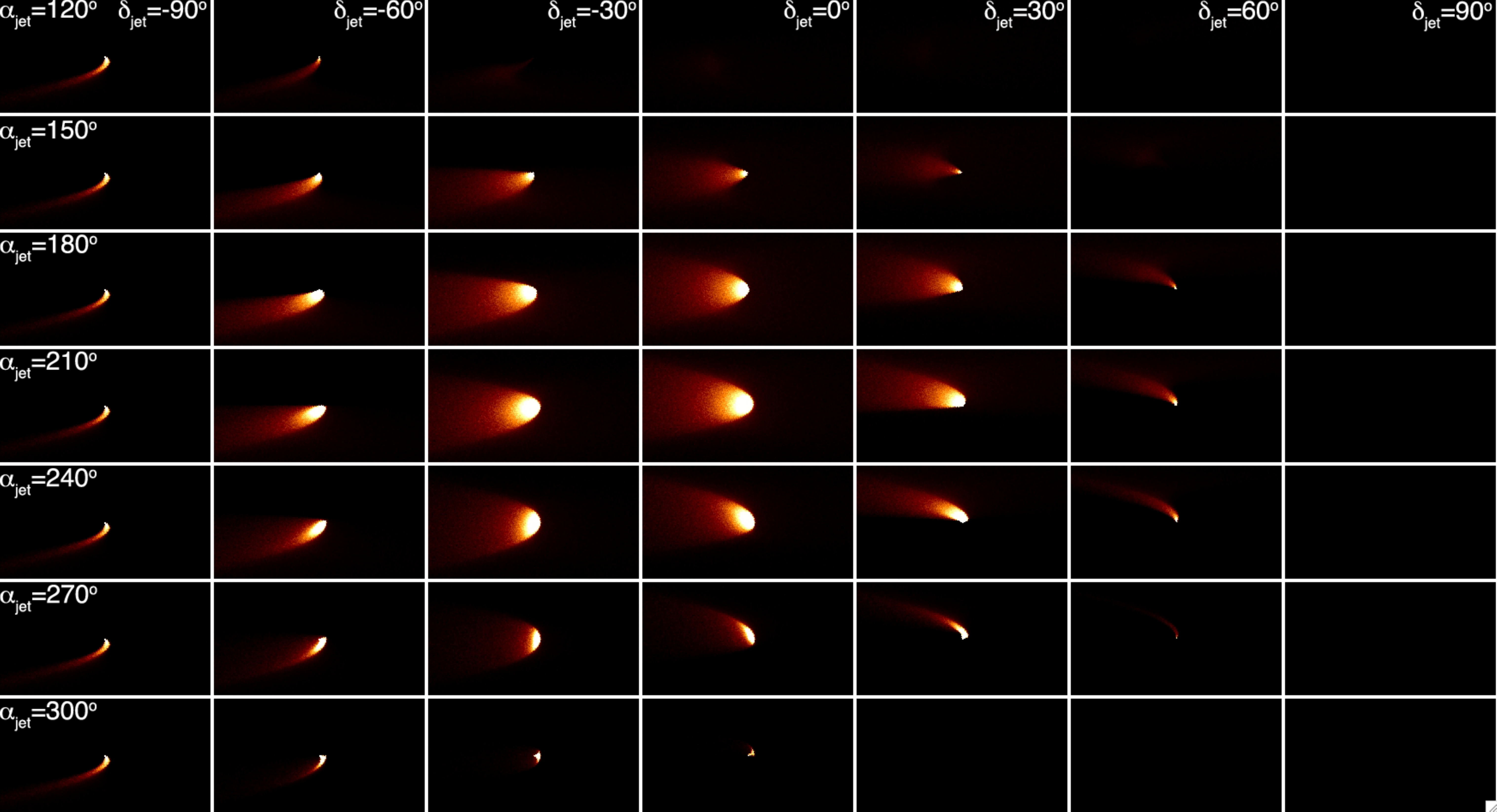}
\caption{\small Images of dust ejection models for 2005 Dec 29 for $10^{-3}<\beta<10^{-1}$ and
different jet directions as labeled, where $\alpha_{\rm jet}$ is constant for each row of models
and $\delta_{\rm jet}$ is constant for each column of models.
In all panels, the source of emission ({\it i.e.}, the nucleus) is at the center of each image.
}
\label{model_20051229}
\end{figure}

We show resulting modeled dust clouds for 2005 November 26 and 2005 December 29 for different jet orientations in
Figures~\ref{model_20051126} and \ref{model_20051229}, respectively.  Comparing these to the observations in Figure~\ref{model_gemini},
we see that reasonable matches to our data are provided by models with jet directions of $180\degr\lesssim\alpha_{\rm jet}\lesssim120$\degr and
$\delta_{\rm jet}\approx-60$\degr, or approximately oriented towards the Sun during the comet's active period.
Crucially, all model scenarios also reproduce the non-detection of activity on 2006 February 03 (Fig.~\ref{images_re70}), where all
dust particles in each scenario have dispersed from the field of view by that time.  This rapid disappearance of dust activity
did not occur from a previous set of test scenarios for which we assumed much larger particles were present ($\beta_{min}=10^{-3}$).
We thus conclude that the dust particles ejected by 176P must have been on the order of $\sim$10~$\mu$m in size
(corresponding to $\beta\sim10^{-1}$), similar to the dominant particle sizes found for 133P \citep{hsi04}.  Using Equation~\ref{ejecvel},
we find an approximate ejection velocity for these particles of $v_{ej}\sim5$~m~s$^{-1}$, or somewhat faster than the $v_{ej}\sim1$~m~s$^{-1}$
ejection velocities found for 133P.

To determine an approximate dust production rate, we refer back to our estimate for 176P's unresolved coma mass in
\S\ref{photactive} of $M_d\sim7\times10^4$~kg.  This mass estimate is based on nucleus photometry employing $3\farcs0$-radius photometry
apertures, which at the distance of 176P in December 2005, correspond to physical radii of $r_{coma}\approx5000$~km.
The time that it would take a dust particle to cross
the photometry aperture can then be estimated as $t_{cross}=r_{coma}/v_{ej}\sim1\times10^6$~s, or about 12 days.  Using
this as the approximate timescale on which dust in the coma must be replaced by new dust production from the nucleus,
we therefore find an approximate dust production rate of $dM/dt\approx M_d/t_{cross}\sim0.07$~kg~s$^{-1}$.  This
production rate is similar to that found for 133P by \citep{hsi04}, but is also likewise probably accurate, at best, to
an order of magnitude.

\begin{figure}
\plotone{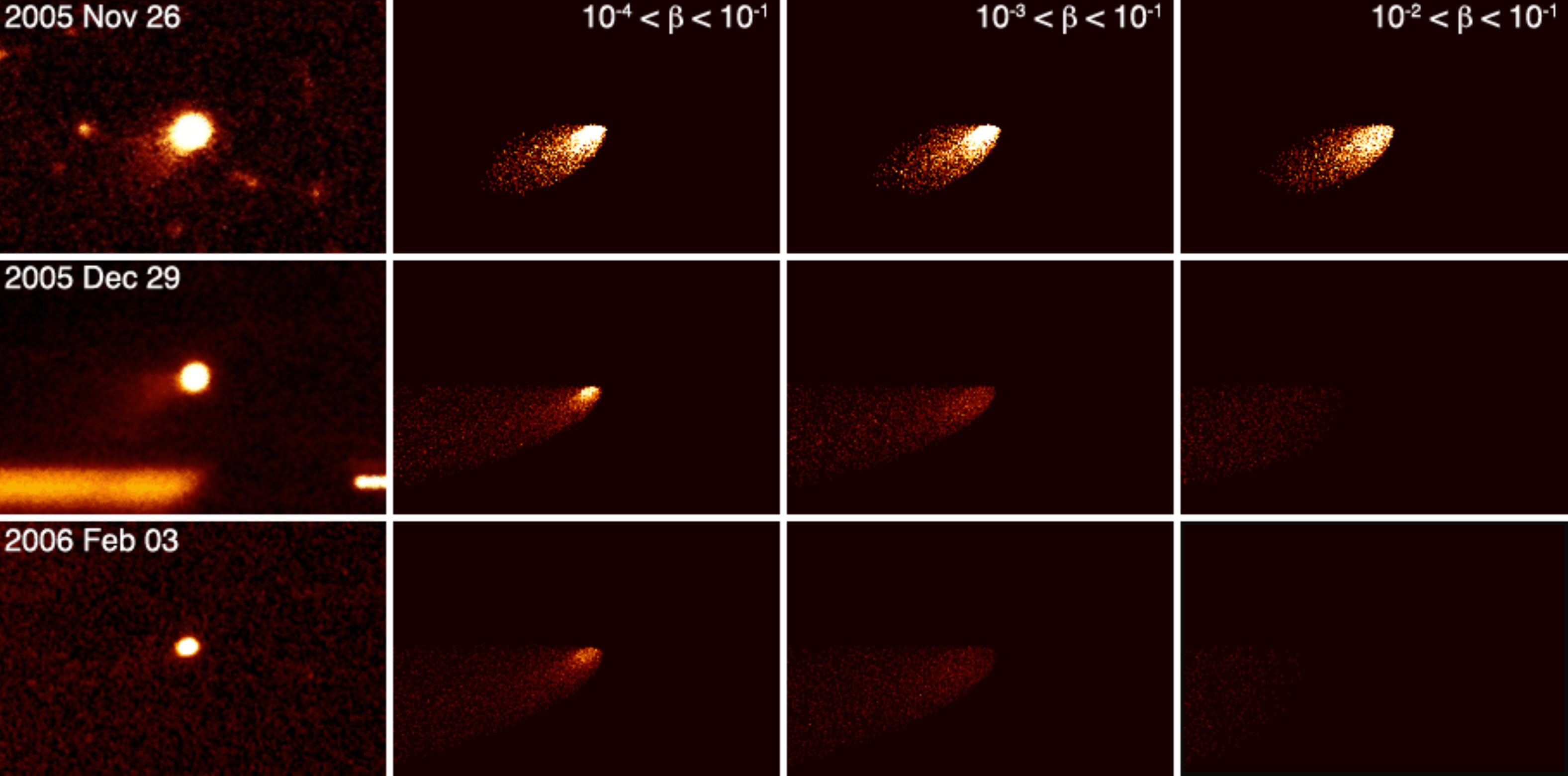}
\caption{\small Comparison of observations (first column of panels) with images of impulsive dust ejection models using different particle
size distributions ($10^{-4}<\beta<10^{-1}$ in the second column, $10^{-3}<\beta<10^{-1}$ in the third column, and $10^{-2}<\beta<10^{-1}$ in
the fourth column).  Data and models for 2005 November 26, 2005 December 29, and 2006 February 03 are shown along the first, second, and
third rows, respectively.  All models consist of dust ejected on a single day on November 15 (four weeks after perihelion and 11 days before
cometary activity was first observed on November 26).
In all panels, the source of emission ({\it i.e.}, the nucleus) is at the center of each image.
}
\label{model_impulse}
\end{figure}

We are further interested in whether 176P's dust emission is only consistent with continuous emission, {\it i.e.}, that
would be expected if the activity is driven by the sublimation of ice, or whether it can also be reproduced by an impulsive emission
event, {\it i.e.}, if 176P's dust tail solely consists of ejecta produced in an impact by another asteroid.  To test this,
we produce another series of models in which dust emission is limited to a single burst of particles.  We test three different
size distributions --- $10^{-4}<\beta<10^{-1}$, $10^{-3}<\beta<10^{-1}$, $10^{-2}<\beta<10^{-1}$ --- where dust in each model is ejected
on a single day (2005 November 15), which is four weeks after perihelion and 11 days before activity was first observed
(Fig.~\ref{model_impulse}).
We then continue to follow the evolution of the dust and compare its appearance in each model to observations on 2005 December 29
and 2006 February 03.

We clearly see that for the models including large particles ($\beta_{min}=10^{-4}$ and $\beta_{min}=10^{-3}$;
second and third columns of Fig.~\ref{model_impulse}), the appearance of the comet remains approximately constant
between November and December, as is observed.  However, the comet retains its appearance through
February as well (consistent with our continuous emission models),
which is inconsistent with observations.  If only smaller particles are included ($\beta_{min}=10^{-2}$;
fourth column of Fig.~\ref{model_impulse}), we note that the dust cloud does dissipate appreciably by February,
but also undergoes significant dissipation by December, inconsistent
with observations.  Thus, we find that a single burst of particles cannot produce a dust cloud that simultaneously
remains constant between 2005 November 26 and 2005 December 29, but then dissipates by 2006 February 03.

We therefore conclude that 176P's activity is most likely driven by
an extended emission event, in which 176P's dust tail is continuously replenished by small, fast-dissipating particles between
2005 November and December, allowing it to maintain its appearance, but where dust emission ceases sometime in December, leading to the nearly
complete dissipation of the dust cloud by the time the comet is observed again on 2006 February 03.  This behavior
strongly suggests that 176P's activity is driven by the sublimation of volatile ices, as was previously
found for fellow MBCs, 133P and 238P \citep{hsi04,hsi09c}.  If this conclusion is correct, we would also expect 176P to exhibit
repeated activity, similar to 133P and 238P \citep{hsi10,hsi10b}, perhaps during its upcoming perihelion passage (2011 July 01).
Observations of renewed activity at this time will provide strong support for sublimation at the cause of 176P's activity.
Conversely, however,
if no activity is observed, reconsideration of impact-generated dust emission as the source of 176P's activity may be necessary.

In terms of pole orientation, our dust modeling analysis is remarkably consistent with the results of our lightcurve analysis
(\S\ref{rotimplications}), as a jet orientation of
$180\degr\lesssim\alpha_{\rm jet}\lesssim120\degr$ and $\delta_{\rm jet}\approx-60\degr$ that is effectively coincident with 
the object's rotational pole corresponds to an orbital obliquity of $50\degr\lesssim\varepsilon\lesssim75\degr$.  The consistency
of these results is encouraging support for their accuracy, and strongly suggests that 176P is in fact a high-obliquity object
that is at solstice around the time that a localized site in the ``summer-time'' hemisphere experiences active sublimation,
just as described under the seasonal heating hypothesis for MBCs.

\section{DISCUSSION}

\subsection{Activity Profile\label{actvprofile}}

Little is currently known about the activity profile of 176P.
During 176P's 2005 active period,
dust emission appeared to end between $\nu\sim19\degr$ and $\nu\sim28\degr$ in true anomaly.  No
conclusive evidence of activity was observed from Lulin Observatory when the comet was at a true
anomaly of $\nu=1.4\degr$, though a photometric analysis
(\S\ref{photactive}) suggests that activity may have been present but simply escaped detection, perhaps due to poor seeing.
As such, the point at which 176P's activity began is essentially
unconstrained.  If 176P is similar to 133P and exhibits recurrent, seasonally-modulated dust emission
that persists over $\sim90\degr$ of its orbit, though, we might have expected it to resume cometary activity
as early as 2010 September when it reached $\nu\sim290\degr$.  If, however, 176P is active over a
much smaller portion of its orbit than 133P (as it might, for example, if shadowing due to local
topography also has a strong activity-modulating effect), then activity may not return until closer
to the object's perihelion passage on 2011 July 01.  Continued monitoring of 176P is highly encouraged
to help clarify this issue.

\subsection{Comparison to other MBCs}

The activity of 176P is similar to 133P \citep{hsi04,hsi10} in terms of dust output ({\it c.f}.\ \S\ref{modeling}), but much weaker
than highly active MBCs like 238P \citep{hsi09c,hsi11}, P/2008 R1 (Garradd) \citep[hereafter P/Garradd;][]{jew09},
or P/2010 R2 (La Sagra) (hereafter P/La Sagra).
Recently observed comet-like activity for two other objects in the main asteroid belt 
--- P/2010 A2 (LINEAR) and (596) Scheila \citep{bir10,lar10} --- is likely to be due to impact-generated ejecta clouds, and not
sublimation-driven dust emission \citep{jew10,sno10,jeh11}, and so we omit these so-called ``disrupted asteroids'' from this discussion.

Differences in MBC activity strength could be due to low-activity MBCs actually being intrinsically
less icy than highly active MBCs, or perhaps simply activated less recently than highly active MBCs,
resulting in greater depletion of their current active sites.  Observations of highly active
MBCs during successive future active periods to track their decline in strength, if any, will help to
clarify this issue.  Noting the distribution of activity levels of MBCs discovered in the future may
also be informative. A high ratio of low-activity objects to highly active objects could indicate
asymptotic declines in activity strength as individual active sites become depleted. On the other hand,
a flatter distribution of MBCs of varying strengths, or evidence that activity strength has a particular
spatial dependence, could indicate that activity strength is more strongly dependent on the intrinsic ice
content of each body (and by extension, the ice content of the surrounding population).
Any analysis of this nature will also have to account for observational
discovery biases towards more highly active objects.

The activity profile discussed above (\S\ref{actvprofile}) is particularly interesting
in relation to 133P in that while 133P becomes active shortly before
perihelion, 176P becomes inactive shortly after perihelion (Fig.~\ref{actv176p}).
This fact is significant because if 176P's thermal inertia
is low, it should reach its maximum overall temperature close to perihelion.
If thermal inertia is high, 176P's temperature may not be at its maximum at
perihelion, but should instead be increasing at this point of close approach
to the Sun.  In either case, the fact that 176P's activity ends (or at least declines
below detectable levels) shortly after the object's overall temperature has peaked or is even
still increasing strongly favors a scenario in which activity is modulated by seasonal effects,
as proposed for 133P \citep{hsi04,hsi10}, rather than proximity to the Sun and overall temperature.

One caveat, though, is that we do not actually know whether the duration of 176P's activity is
similar to 133P's ({\it i.e.}, persisting over $\sim$25\% of its orbit).
The active portion of 176P's orbit might still be centered on perihelion if, in
addition to ending just after perihelion, it also only starts just before perihelion, giving
it a much shorter active arc than 133P.  Such a shorter active arc could perhaps be due to a much
smaller (or older and therefore more depleted) supply of exposed volatile material, or topography
imparting much more local shadowing, and therefore a much sharper seasonal effect, than on 133P.  Again, as discussed above, continued
monitoring of the object in search of resumed activity as it approaches its next perihelion passage
will be essential for resolving these questions.

\section{SUMMARY}

We present a physical analysis of main-belt comet 176P/LINEAR based on optical observations
and numerical modeling, finding the following key results:

\begin{enumerate}
\item{In optical imaging data obtained between late 2005 and mid-2009, we detect
the presence of cometary activity for 176P between 2005 November 26 and 2005 December 29,
and do not detect activity between 2006 February 03 and 2009 May 03, placing the apparent
cessation point for 176P's activity at 20$^{\circ}$-25$^{\circ}$ past perihelion.
The turn-on point for activity remains unknown at this time.}
\item{Using photometric data obtained when no activity was detected for 176P, we find
best-fit IAU phase function parameter values of $H=15.10\pm0.05$~mag and $G=0.15\pm0.10$,
and best-fit linear phase function parameter values of $m(1,1,0)=15.35\pm0.05$~mag and
$\beta=0.038\pm0.005$~mag~deg$^{-1}$.}
\item{Using data obtained in 2007 when no activity was detected, we also find a rotation
period solution of $P_{rot}=22.23\pm0.01$~hr and a photometric range of $\Delta m\sim0.7$~mag.
Using this rotational period to phase data from 2005 when 176P was active, we find a much smaller
photometric range of $\Delta m\sim0.2$~mag which cannot be fully accounted for by coma contamination
and as such is attributed by us to viewing geometry effects.  From these lightcurve measurements, we
derive a likely orbital obliquity of $\varepsilon\sim60^{\circ}$ and an axis ratio of $1.8<b/a<2.1$.}
\item{In performing dust modeling of 176P's activity, we noted that the orientation of the dust tail does not coincide
with any syndyne or synchrone curves, requiring us to consider asymmetric dust emission, {\it e.g.},
a cometary jet.  We find plausible fits to our data using a continuous dust emission model
with a particle size range of $9\times10^{-2}<\beta<10^{-1}$ and a reference ejection velocity of $v_0=60$~m~s$^{-1}$.
Our data are not consistent with any impulsive dust emission scenarios that we investigated.
We find a likely jet orientation (and therefore rotational pole orientation)
of $180\degr\lesssim\alpha_{\rm jet}\lesssim120$\degr and $\delta_{jet}\approx-60^{\circ}$, in good
agreement with our lightcurve analysis above.}
\end{enumerate}

\section*{Acknowledgements}
We appreciate support for HHH by the National Aeronautics and Space Administration (NASA)
through Hubble Fellowship grant HF-51274.01, awarded by the Space Telescope Science Institute,
which is operated by the Association of Universities for Research in Astronomy, Inc., for NASA,
under contract NAS 5-26555, and by the United Kingdom's Science and Technology Facilities Council (STFC)
through STFC fellowship grant ST/F011016/1.
Additional support was provided by NASA through a planetary astronomy grant to DJ, and by a Royal
Society Newton Fellowship grant and Michael West Fellowship grant to PL.
We thank Richard Wainscoat, Dale Kocevski, Jana Pittichov\'a, Rita
Mann, and Colin Snodgrass for donated telescope time,
National Central University in Taiwan for access to Lulin Observatory,
and the STFC, through the Panel for the Allocation of Telescope Time (PATT), and
the European Southern Observatory (ESO) for travel support.
We are also grateful to Ian Renaud-Kim, Dave Brennen, and John Dvorak at the UH 2.2~m,
Kathy Roth, Chad Trujillo, and Tony Matulonis at Gemini, Greg
Wirth, Cynthia Wilburn, and Gary Punawai at Keck, Ming-Hsin Chang at Lulin,
Dave Jones at the INT, Antonio Garcia at the WHT, and Leonardo Gallegos at the NTT
for their assistance with our observations,
and to an anonymous referee for helpful comments on this manuscript.

\begin{deluxetable}{llrrcccrcrrrcc}
\scriptsize
\tablewidth{0pt}
\tablecaption{Observation Log\label{obslog}}
\tablecolumns{14}
\tablehead{
  \colhead{UT Date}
   & \colhead{Tel.\tablenotemark{a}}
   & \colhead{$N$\tablenotemark{b}}
   & \colhead{$t$\tablenotemark{c}}
   & \colhead{Filters}
   & \colhead{$\theta_s$\tablenotemark{d}}
   & \colhead{Active?}
   & \colhead{$\nu$\tablenotemark{e}}
   & \colhead{$R$\tablenotemark{f}}
   & \colhead{$\Delta$\tablenotemark{g}}
   & \colhead{$\alpha$\tablenotemark{h}}
   & \colhead{$\alpha_{pl}$\tablenotemark{i}}
   & \colhead{$m_R$\tablenotemark{j}}
   & \colhead{$m_R(1,1,\alpha)$\tablenotemark{k}} \\
}
\startdata
2005 Oct 18 & \multicolumn{6}{l}{{\it Perihelion}
  ........................................................... }
   & 0.0 & 2.58 & 1.59 & 1.5 & --0.1 & \multicolumn{2}{l}{ ............................................ } \\
2005 Oct 24 & Lulin  & 10 &  3000 & $R$        & 1.5 & yes? &   1.4 & 2.58 & 1.60 &  4.3 & --0.1 & 18.20$\pm$0.01 & 15.12$\pm$0.01 \\
2005 Nov 26 & Gemini &  2 &   240 & $r'$       & 0.6 & yes &  10.1 & 2.59 & 1.82 & 16.4 & --0.1 & 19.11$\pm$0.04 & 15.74$\pm$0.04 \\
2005 Dec 22 & UH/Tek &  7 &  2100 & $R$        & 1.1 & yes &  16.8 & 2.60 & 2.12 & 21.1 & --0.1 & 19.65$\pm$0.01 & 15.95$\pm$0.01 \\
2005 Dec 24 & UH/Tek & 26 &  7800 & $R$        & 1.0 & yes &  17.3 & 2.60 & 2.15 & 21.3 & --0.1 & 19.46$\pm$0.01 & 15.73$\pm$0.01 \\
2005 Dec 25 & UH/Tek & 33 &  9900 & $BVRI$     & 0.9 & yes &  17.6 & 2.60 & 2.16 & 21.4 & --0.1 & 19.62$\pm$0.01 & 15.88$\pm$0.01 \\
2005 Dec 26 & UH/Tek & 31 &  9300 & $BVRI$     & 1.0 & yes &  17.8 & 2.60 & 2.17 & 21.5 & --0.1 & 19.62$\pm$0.01 & 15.86$\pm$0.01 \\
2005 Dec 27 & UH/Tek & 29 &  8700 & $R$        & 0.8 & yes &  18.1 & 2.60 & 2.19 & 21.5 & --0.1 & 19.59$\pm$0.01 & 15.81$\pm$0.01 \\
2005 Dec 29 & Gemini & 34 &  3060 & $g'r'i'z'$ & 0.7 & yes &  18.6 & 2.60 & 2.21 & 21.7 & --0.1 & 19.62$\pm$0.01 & 15.80$\pm$0.01 \\
2006 Feb 03 & UH/Tek & 11 &  3300 & $R$        & 1.1 &  no &  27.7 & 2.63 & 2.71 & 21.2 &   0.0 & 20.25$\pm$0.01 & 15.98$\pm$0.01 \\ 
2006 Aug 31 & UH/Tek &  2 &   600 & $R$        & 0.8 &  no &  75.1 & 2.93 & 3.55 & 14.3 &   0.0 & 21.23$\pm$0.08 & 16.14$\pm$0.08 \\ 
2006 Sep 02 & UH/Tek &  2 &   600 & $R$        & 0.9 &  no &  75.5 & 2.94 & 3.53 & 14.6 &   0.0 & 21.09$\pm$0.05 & 16.00$\pm$0.05 \\ 
2006 Dec 11 & UH/Opt & 30 &  9000 & $R$        & 1.0 &  no &  94.4 & 3.12 & 2.42 & 14.5 & --0.1 & 20.01$\pm$0.01 & 15.62$\pm$0.01 \\
2006 Dec 16 & UH/Opt & 11 &  3300 & $R$        & 0.9 &  no &  95.3 & 3.13 & 2.38 & 13.3 & --0.1 & 20.13$\pm$0.01 & 15.77$\pm$0.01 \\ 
2006 Dec 18 & UH/Opt &  3 &   900 & $R$        & 1.3 &  no &  95.6 & 3.14 & 2.36 & 12.8 & --0.1 & 20.09$\pm$0.09 & 15.74$\pm$0.09 \\ 
2007 Jan 27 & Keck   &  3 &   240 & $R$        & 0.9 &  no & 102.5 & 3.21 & 2.23 &  0.8 & --0.1 & 19.50$\pm$0.01 & 15.23$\pm$0.01 \\
2007 Feb 15 & UH/Tek & 30 &  9000 & $R$        & 0.9 &  no & 105.6 & 3.25 & 2.33 &  7.5 &   0.0 & 19.87$\pm$0.01 & 15.47$\pm$0.01 \\
2007 Feb 16 & UH/Tek & 49 & 14700 & $R$        & 1.2 &  no & 105.8 & 3.25 & 2.33 &  7.8 &   0.0 & 19.93$\pm$0.01 & 15.53$\pm$0.01 \\
2007 Mar 21 & UH/Tek & 35 & 10500 & $R$        & 0.8 &  no & 111.1 & 3.31 & 2.72 & 15.4 &   0.0 & 20.71$\pm$0.01 & 15.94$\pm$0.01 \\
2007 Mar 22 & UH/Tek & 51 & 15300 & $R$        & 1.3 &  no & 111.3 & 3.31 & 2.73 & 15.5 &   0.0 & 20.82$\pm$0.01 & 16.04$\pm$0.01 \\
2007 May 19 & UH/Tek &  7 &  2100 & $R$        & 1.1 &  no & 120.2 & 3.41 & 3.64 & 16.1 &   0.1 & 21.57$\pm$0.05 & 16.10$\pm$0.05 \\
2008 Jun 29 & NTT    &  2 &   360 & $R$        & 1.3 &  no & 173.2 & 3.80 & 3.80 & 15.4 &   0.1 & 21.68$\pm$0.07 & 15.88$\pm$0.07 \\ 
2008 Jun 30 & NTT    &  3 &   540 & $R$        & 1.0 &  no & 173.3 & 3.80 & 3.81 & 15.3 &   0.1 & 21.70$\pm$0.05 & 15.90$\pm$0.05 \\
2008 Jul 01 & NTT    &  3 &   540 & $R$        & 1.1 &  no & 173.4 & 3.80 & 3.83 & 15.3 &   0.1 & 21.63$\pm$0.05 & 15.81$\pm$0.05 \\
2008 Aug 25 & \multicolumn{6}{l}{{\it Aphelion}
  ............................................................. }
   & 180.0 & 3.81 & 4.52 & 9.9 & 0.0 & \multicolumn{2}{l}{ ............................................ } \\
2009 Jan 23 & WHT    &  4 &   240 & $R$        & 0.8 &  no & 198.0 & 3.77 & 4.01 & 14.1 &   0.0 & 21.47$\pm$0.10 & 15.57$\pm$0.10 \\
2009 May 03 & INT    &  2 &   600 & $R$        & 1.8 &  no & 210.5 & 3.69 & 2.70 &  3.8 &   0.1 & 20.30$\pm$0.04 & 15.31$\pm$0.04 \\
2011 Jul 01 & \multicolumn{6}{l}{{\it Perihelion}
  ........................................................... }
   & 0.0 & 2.58 & 2.98 & 19.4 & 0.1 & \multicolumn{2}{l}{ ............................................ }
\enddata
\tablenotetext{a}{Telescope used}
\tablenotetext{b}{Number of images}
\tablenotetext{c}{Total effective exposure time}
\tablenotetext{d}{FWHM seeing in arcsec}
\tablenotetext{e}{True anomaly in degrees}
\tablenotetext{f}{Heliocentric distance in AU}
\tablenotetext{g}{Geocentric distance in AU}
\tablenotetext{h}{Solar phase angle (Sun-176P-Earth) in degrees}
\tablenotetext{i}{Orbit plane angle (between the observer and object orbit
                    plane as seen from the object) in degrees}
\tablenotetext{j}{Mean (in magnitude space) of maximum and minimum $R$-band magnitudes measured for nucleus}
\tablenotetext{k}{Inferred reduced $R$-band magnitude (normalized to $R=\Delta=1$~AU)
      at midpoint of full photometric range (assumed to be 0.70 mag) of rotational lightcurve}
\end{deluxetable}

\begin{deluxetable}{lccc}
\tablewidth{0pt}
\tablecaption{Color Measurements\label{colors}}
\tablehead{
  \colhead{UT Date}
   & \colhead{$B-V$}
   & \colhead{$V-R$}
   & \colhead{$R-I$}
}
\startdata
2005 Dec 25 & 0.61$\pm$0.03 & 0.32$\pm$0.02 & 0.27$\pm$0.03 \\
2005 Dec 25 & 0.61$\pm$0.03 & 0.36$\pm$0.03 & 0.27$\pm$0.04 \\
2005 Dec 26 & 0.64$\pm$0.03 & 0.37$\pm$0.02 & 0.35$\pm$0.03 \\
2005 Dec 29 & 0.64$\pm$0.02 & 0.36$\pm$0.02 & 0.32$\pm$0.02 \\
\hline
Mean        & 0.63$\pm$0.02 & 0.35$\pm$0.02 & 0.31$\pm$0.04 
\enddata
\end{deluxetable}

\begin{deluxetable}{lccccc}
\tablewidth{0pt}
\tablecaption{Lightcurve Observations\label{ltcobs}}
\tablecolumns{6}
\tablehead{
  \colhead{UT Date}
   & \colhead{$\Delta t_{obs}$\tablenotemark{a}}
   & \colhead{Range\tablenotemark{b}}
   & \colhead{$m_{mid}$\tablenotemark{c}}
   & \colhead{$m_{mid}(1,1,\alpha)$\tablenotemark{d}}
}
\startdata
2006 Feb 03 & 1.65 & 0.13$\pm$0.06 & 20.28$\pm$0.06 & 16.02$\pm$0.29 \\ 
2006 Dec 11 & 4.23 & 0.39$\pm$0.12 & 20.02$\pm$0.12 & 15.63$\pm$0.20 \\ 
2006 Dec 16 & 1.98 & 0.28$\pm$0.05 & 20.10$\pm$0.05 & 15.74$\pm$0.22 \\ 
2007 Feb 15 & 6.71 & 0.51$\pm$0.04 & 19.92$\pm$0.04 & 15.62$\pm$0.04 \\ 
2007 Feb 16 & 7.34 & 0.55$\pm$0.06 & 20.01$\pm$0.06 & 15.68$\pm$0.06 \\ 
2007 Mar 21 & 4.66 & 0.63$\pm$0.07 & 20.77$\pm$0.07 & 16.00$\pm$0.08 \\ 
2007 Mar 22 & 5.12 & 0.55$\pm$0.10 & 20.83$\pm$0.10 & 16.05$\pm$0.13 \\ 
2007 May 19 & 0.86 & 0.31$\pm$0.22 & 21.53$\pm$0.22 & 16.06$\pm$0.32    
\enddata
\tablenotetext{a}{Time spanned by observations (hr)}
\tablenotetext{b}{Photometric range between maximum and minimum $R$-band magnitudes measured for the nucleus}
\tablenotetext{c}{Midpoint between maximum and minimum $R$-band magnitudes measured for the nucleus}
\tablenotetext{d}{Inferred reduced $R$-band magnitude (normalized to $R=\Delta=1$~AU)
      at midpoint of full photometric range (assumed to be 0.70 mag) of rotational lightcurve}
\end{deluxetable}

\end{document}